\documentclass{aa}
\usepackage{graphicx}
\usepackage{amssymb,amsmath}
\usepackage{mathtools}
\usepackage{natbib}
\usepackage{aas_macros}
\usepackage{subfigure}
\def\void#1{{}}

\newcommand{\bmath}[1]{\mbox{ \boldmath $\!#1\!$ \unboldmath}}

\newcommand{\txn}[1]{\textnormal{#1}}

\newcommand{\mJy}{\hbox{\,$\mu$Jy}}
\newcommand{\micron}{\hbox{\,$\mu$m}}

\newcommand{\irx}{\hbox{$\langle IRX \rangle$}}

\newcommand{\Mstar}{\hbox{{$\txn{M}_{\ast}$}}}
\newcommand{\Msun}{\hbox{$\txn{M}_{\odot}$}}

\newcommand{\MsunYr}{\hbox{$\txn{M}_{\odot}\txn{yr}^{-1}$}}
\newcommand{\invYr}{\hbox{$\txn{yr}^{-1}$}}

\newcommand{\SFR}{\hbox{$\txn{SFR}$}}
\newcommand{\sSFR}{\hbox{$\txn{sSFR}$}}

\newcommand{\SFRsed}{\hbox{$\txn{SFR}^\txn{SED}$}}
\newcommand{\SFRnrk}{\hbox{$\txn{SFR}^\txn{NRK}$}}

\newcommand{\sSFRsed}{\hbox{$\txn{sSFR}^\txn{SED}$}}
\newcommand{\sSFRnrk}{\hbox{$\txn{sSFR}^\txn{NRK}$}}

\newcommand{\Lir}{\hbox{$L_\txn{IR}$}}
\newcommand{\LirSED}{\hbox{$L_\txn{IR}^\txn{SED}$}}
\newcommand{\LirNRK}{\hbox{$L_\txn{IR}^\txn{NRK}$}}
\newcommand{\LirMU}{\hbox{$L_\txn{IR}^{24\mu}$}}
\newcommand{\Luv}{\hbox{$L_\txn{UV}$}}

\newcommand{\taubP}{\hbox{$\tau_{B\perp}$}}
\newcommand{\tauLbc}{\hbox{$\hat{\tau}^\txn{BC}_\lambda$}}
\newcommand{\tauLTh}{\hbox{$\hat{\tau}_\lambda(\theta)$}}
\newcommand{\tauVbc}{\hbox{$\hat{\tau}^\txn{BC}_V$}}
\newcommand{\tauLismTh}{\hbox{$\hat{\tau}^\txn{ISM}_\lambda(\theta)$}}
\newcommand{\meantauVISM}{\hbox{$\langle\hat{\tau}^\txn{ISM}_V\rangle_\theta$}}
\newcommand{\meantaulISM}{\hbox{$\langle\hat{\tau}^\txn{ISM}_\lambda\rangle_\theta$}}

\newcommand{\tauVyoung}{\hbox{$\hat{\tau}^\txn{young}_V(\theta)$}}
\newcommand{\tauVold}{\hbox{$\hat{\tau}^\txn{old}_V(\theta)$}}
\newcommand{\nVism}{\hbox{$n^\txn{ISM}_V(\theta)$}}

\newcommand{\NRKv}{\hbox{$\bmath{NRK}$}}
\newcommand{\NRK}{\hbox{$NRK$}}
\newcommand{\NUVrK}{\hbox{$\txn{NUVrK}$}}
\begin{document}
\title{Encoding of the infrared excess  (IRX) in the  \NUVrK\ color diagram for star-forming galaxies % I :  The COSMOS 24$\mu$m sample
 % a single parameter combining the luminosities $L_{NUV}, L_r, L_K$. 
%\thanks{ }
}
\author{  S. Arnouts \inst{1,4}, E. Le~Floc'h\inst{2}, J. Chevallard\inst{3}, B. D. Johnson\inst{3}, O. Ilbert\inst{4}, M. Treyer\inst{4}, H. Aussel\inst{2}, P. Capak\inst{5},  D. B. Sanders\inst{6},  N. Scoville\inst{7},  H. J. McCracken\inst{3},  B. Milliard\inst{4}, L. Pozzetti\inst{8}, M. Salvato\inst{9}  }
\offprints{St\'ephane Arnouts,  \email{arnouts@cfht.hawaii.edu}  }
\institute{ 
 Canada-France-Hawaii Telescope Corporation, 65-1238 Mamalahoa Hwy,  Kamuela, HI  96743, USA 
  \and
  Laboratoire AIM, CEA/DSM-CNRS-Universit\'e Paris Diderot, IRFU/Service dÕAstrophysique, B\^at. 709, CEA-Saclay, 91191 Gif- sur-Yvette Cedex, France
 \and
  UPMC-CNRS, UMR7095, Institut d'Astrophysique de Paris, F-75014, Paris, France
 \and
  Aix Marseille Universit\'e, CNRS, Laboratoire d'Astrophysique de Marseille, UMR 7326, 13388, Marseille, France
    \and
    Spitzer Science Center, MS 314-6, California Institute of Technology, Pasadena, CA 91125
    \and
    Institute for Astronomy, 2680 Woodlawn Dr., University of Hawaii, Honolulu, Hawaii, 96822
   \and
    California Institute of Technology, MC 249-17, 1200 East California Boulevard, Pasadena, CA 91125
     \and
     INAF - Osservatorio Astronomico di Bologna - Via Ranzani 1 - I-40127 Bologna - Italy
     \and
     MPE, Giessenbachstrasse 1, D-85748, Germany; Excellence Cluster , Boltzmann Strasse 2, D-85748, Germany
    	  }
\date{Received 25 April 2013. Accepted 25 July 2013.}
\titlerunning{Encoding of the IRX in  \NUVrK\  diagram}
\authorrunning{Arnouts et al. }
%-------------------------------
\abstract{
 We present an empirical method of assessing the star formation rate (SFR) of star-forming galaxies based on their locations in the rest-frame color-color diagram $(NUV-r)$ vs $(r-K)$. 
 By using the Spitzer 24\,\micron\ sample in the COSMOS field ($\sim 16\,400$ galaxies with $0.2\le z\le 1.3$) and a local GALEX-SDSS-SWIRE sample ($\sim700$ galaxies with $z\le 0.2$), we show that the mean infrared excess \irx$=\langle \Lir/\Luv \rangle$ can be described by a single vector, \NRKv, that combines the two colors. The calibration between \irx\ and NRK allows us to recover the IR luminosity, $\Lir$, with an accuracy of $\sigma \sim$0.21 for the COSMOS sample  and 0.27 dex for the local one. The SFRs derived with this method agree with the ones based on the observed (UV+IR) luminosities and on the spectral energy distribution fitting for the vast majority ($\sim$ 85\%) of the star-forming population.
 Thanks to a library of model galaxy SEDs with realistic prescriptions for the star formation history, we show that we need to include a two-component dust model (i.e., birth clouds and diffuse ISM) and a full distribution of  galaxy inclinations in order to reproduce the behavior of the \irx\ stripes in the  \NUVrK\ diagram.
 In conclusion, the NRK method, based only on the rest-frame UV/optical colors available in most of the extragalactic fields, offers a simple alternative of assessing the SFR of star-forming galaxies in the absence of far-IR or spectral diagnostic observations. 
}
\maketitle

\section{Introduction}
Star formation activity is a key observable for understanding the physical processes in the build up of galaxies.
 The SFR in galaxies  depends  on the physics of star-forming regions, merger history, gas infall and outflows and on stellar and AGN feedbacks. The SFR distribution and its evolution therefore offer a crucial test for any model of galaxy evolution. There are different indicators of the ongoing star formation \citep[e.g.,][]{Kennicutt1998}, such as ultraviolet continuum ($\lambda\sim 912-3000$ \AA) produced by massive, young stars ($t\sim10^8$ yr, \citealt{Martin2005}); nebular recombination lines from gas ionized by the hot radiation from early-type stars ($\lambda\le 912$\,\AA, $t\sim10^7$ yr); far-infrared (FIR) emission from dust heated by UV light;  and non-thermal radio emission, such as synchrotron radiation from supernova remnants (see \citealt{Kennicutt1983, Kennicutt1998} for their respective calibrations with SFR). 
 
 Recently, measurements of SFRs in large samples of galaxies based on the above indicators have provided interesting insights in the dominant mode of baryon accretion for star-forming galaxies: the tight correlation, with small scatter, between stellar mass and SFR observed up to $z\sim2$ \citep[e.g.,][]{Salim2007, Noeske2007a, Daddi2007, Wuyts2011b} suggests that a secular, smooth process, such as gas accretion, rather than merger-induced starbursts, may be the dominant mechanism governing star formation in galaxies. Also, the relation between specific SFR and stellar mass \citep[e.g.,][]{Brinchmann2004, Salim2007, Noeske2007b} reveals that less massive galaxies have their onset of star formation occurring later than more massive ones \citep[e.g.,][]{Cowie1996}, and that a simple  mass-dependent gas exhaustion model can reproduce the observed decline of the cosmic SFR since $z\sim 1-1.5$ \citep[e.g.,][]{Lilly1996, Schiminovich2005, Lefloch2005}. Such a scenario is also consistent with recent evidence of a higher fraction of molecular gas in massive star-forming galaxies at $z\ge1$  with respect to nearby galaxies \citep[e.g.,][]{Erb2006, Daddi2008, Daddi2010, Tacconi2010}. 

Extending such measurements to high redshift for large samples of galaxies poses several challenges: optical recombination lines are often too weak and are shifted to near-IR wavelengths, where current spectroscopic capabilities are limited. Far-IR indicators are also of limited use at high redshift, since the modest sensitivity and resolution of infrared telescopes make these observations sparse and restricted to the most massive galaxies, unless stacking techniques are used \citep[e.g.,][]{Zheng2007b, Pannella2009, Karim2011, Lee2010, Heinis2013}. On the other hand, with increasing redshift,  the ultraviolet continuum emission is progressively shifted to optical and near-IR bands, making it easily accessible with the largest ground-based telescopes \citep[e.g.,][]{Burgarella2007, Daddi2007, Reddy2006, Steidel1996, Bouwens2011}. 

In the absence of dust, the ultraviolet luminosity of a galaxy is proportional to the mean SFR on a timescale $t \sim 10^8$ yr \citep{Donas1984, Leitherer1995}. The presence of dust in the interstellar medium of galaxies hampers such a direct estimate of the SFR from UV observations. Starlight, especially in the UV, is efficiently absorbed and scattered by dust grains, which heat up and re-emit the absorbed energy at FIR wavelengths. This means that a fraction of UV photons will not escape the galaxy,  and that neglecting this effect will lead to severe underestimates of the `true' SFR \citep[e.g.,][]{Salim2007}.
\citet{Calzetti1997} and \citet{Meurer1999} proposed a way to correct the UV continuum for dust attenuation that relies on the existence of a tight correlation between the slope of the UV continuum in the region $1300 \le \lambda / \txn{\AA} \le 2600$ (i.e., $\beta$-slope) and the ratio between the total IR ($8\le \lambda/ \micron \le 1000$) and ultraviolet luminosities (i.e.,  infrared excess, IRX$\sim \Lir/\Luv$). This relation allows the estimation of the total IR luminosity from the shape of the ultraviolet spectral region, providing a way to quantify the amount of attenuation of UV starlight (i.e.,  $A_{UV}$) reprocessed by dust.  
It has been abundantly used to derive the SFR of high redshift galaxies, for which the UV slope is the only accessible quantity \citep{Reddy2006, Smit2012}.

 To account for departures of quiescent star-forming galaxies from the star-burst  IRX--$\beta$ relation, several authors \citep[e.g.,][]{Kong2004, Seibert2005, Salim2007} proposed a modified version of this relation, expressed in terms of $A_{FUV}-\beta$, which leads to a smaller dust correction at a given slope $\beta$, though with a significant scatter [$\sigma(A_{FUV})\sim 0.9$]. It is worth mentioning that such a large scatter is not surprising in view of  the steepness of the $A_{FUV}-\beta$ relation, where a small uncertainty in UV color [e.g., $\Delta (FUV-NUV)\sim 0.1$]  produces a  large variation in $ A_{FUV}$ [($\Delta A_{FUV}\sim 0.8$].
\citet{Johnson2007} proposed another method calibrated for quiescent star-forming galaxies which relies on the combination of the $D_{4000}$ break, a spectral feature sensitive to the age of a galaxy, and a long baseline color. They explored different colors and functions and obtain the smallest residuals by using the $(NUV-3.6 \,\micron)$ color [$\sigma(IRX)\sim0.3$ for their star-forming population]. \citet{Treyer2007} compared the SDSS SFR estimates of \citet{Brinchmann2004}, based on nebular recombination lines, with those obtained from the UV continuum corrected for the effect of dust attenuation with the methods of \citet{Seibert2005, Salim2007}, and of \citet{Johnson2007}. They found that the method of \citet{Johnson2007} leads to the smallest scatter [$\sigma(SFR)\sim0.22$  vs 0.33]. However, the method of \citet{Johnson2007} depends on spectral diagnostics, such as the $D_{4000}$ break, which are difficult to obtain for high redshift galaxies. 

  In this work, we analyze the behavior of the IRX within the rest-frame color-color diagram $(NUV-r)$ vs $(r-K)$. We describe a new relationship  between  IRX  and  a single vector,  called \NRKv, defined as the  combination of the colors $(NUV-r)$ and $(r-K)$. 
 This new diagnostic provides an effective way to assess the total IR luminosity $\Lir$ and then the SFR for individual galaxies with photometric information widely available in large surveys, and does not require complex modeling such as  SED fitting techniques. 
  The $(NUV-r)$ vs $(r-K)$ color diagram adopted in this work is similar to the $(U-V)$ vs $(V-J)$  diagram proposed by \citet{Williams2009}  to separate passive or quiescent from star-forming galaxies, but better leverages the role of SFH and dust by extending into the extreme wavelengths of the SED.  %, which are still available in most extragalactic fields. 
 Total infrared luminosities were estimated from the deep MIPS-24\micron\ observations of the COSMOS field, with standard techniques extrapolating mid-IR  flux densities 
  with luminosity-dependent IR galaxy SED templates as described in e.g.,, \citet{Lefloch2005}. Our approach is motivated by the tight correlations that exist between mid-IR and total IR luminosities, both  at low redshifts \citep{Chary2001, Takeuchi2003} and in the more distant Universe \citep{Bavouzet2008}. In fact, stacking analysis and direct individual identifications of distant sources in the far-IR with facilities like $Spitzer$ and $Herschel$
\citep[e.g.,,][]{Papovich2007,Lee2010,Elbaz2010} have  shown that these extrapolations from mid-IR photometry provide reliable estimates of total IR luminosities up to $z \sim 1.3$ (dispersion of $\sim$ 0.25\,dex, no systematic offset), at least for star-forming galaxies initially selected at 24\micron. 
 Also, given the relative depths of the different IR and sub-millimeter observations carried out in the COSMOS field with e.g.,, $Spitzer$, $Herschel$ or $JCMT$, we note that the deep 24\micron\ data in COSMOS provide the largest sample of  star-forming galaxies 
 with measurable IR luminosities. 
    We  thus decide to limit our analysis to massive ($\txn{M}\ge 10^{9.5}\,\Msun$) star-forming sources first selected at 24\micron\ and lying at $z\le 1.3$ so as to obtain reliable $\Lir$ estimates. In a companion paper \citep{Lefloch2013},  we will extend our analysis to higher redshifts and lower masses by stacking along the \NRKv\ vector the MIPS and Herschel data available in the COSMOS field.  

    The paper is organized as follows. In Section~2 we describe the dataset, the sample selection and the estimates of physical parameters.
  In Section~3 we discuss the behavior of the mean IRX (i.e., \irx) in the $(NUV-r)$ vs $(r-K)$ diagram and the calibration of the \irx\  vs \NRK\  relation. In Section~4, we compare our predicted IR luminosities with the ones based on the 24\,\micron\ observations and SED fitting technique. 
 We also investigate the  dependence with galaxy physical parameters and apply our method to the well established SFR-mass relation. In Section~5 we discuss the method, the possible origin of the relation with a complete library of model galaxy SEDs  and dust models. We draw the conclusions in Section~6.  \\
 Throughout the paper we adopt the following cosmology:  $H_0$=70\  km s$^{-1}$ Mpc$^{-1}$ and $\Omega_M=0.3$, $\Omega_{\Lambda}=0.7$. We adopt  the initial mass function of \citet{Chabrier2003} , truncated at 0.1 and 100 \Msun. All magnitudes are given in the AB system \citep{Oke1974}.

\section{The dataset}
\subsection{Observations}
\subsubsection{The {\it Spitzer}-MIPS 24\,\micron\  observations and COSMOS catalog} 
 The region  of the sky covered by the {\it Cosmic Evolution Survey} \citep[i.e., COSMOS][]{Scoville2007} is observed by {\it Spitzer}--MIPS (Multi Imaging Photometer) at 24, 70 and 160 \micron\ over a 2 deg$^2$ area \citep{Sanders2007}. In this work, we use the deep 24 \micron\ observations and the catalogue of sources extracted by \citet{Lefloch2009}.  This catalog is 90 \% complete down to the flux limit density $S_{24}\sim 80 \, \mJy$. The sources are cross-matched with the multi-wavelength photometric COSMOS catalog. This includes deep ultraviolet GALEX imaging \citep{Zamojski2007}, ground based optical observations with intermediate and broad band filters, near IR photometry \citep{Capak2007, McCracken2010, Taniguchi2007} and deep IRAC photometry \citep{Sanders2007, Ilbert2010} for a total of 31 filter passbands. The combined photometry, as well as accurate photometric redshifts, are available from the catalog of \citet[version 1.8]{Ilbert2010}.  
 The typical depths at $5\sigma$ are 25.2, 26.5, 26.0, 26.0, 25.0, 23.5 and 24.0 in the $NUV,\ u,\ g,\ r,\ i,\ K$ and\ 3.6\,\micron\ pass-bands, respectively. A vast majority of 24\,\micron\ sources is also detected in the optical bands ($\sim 95$ \% with $u\le 26.5$ and $i_\txn{AB}\le 24.5$).

 We adopt spectroscopic redshifts from the bright and faint zCOSMOS sample \citep{Lilly2009}, when available, otherwise we use the photometric redshifts computed by \citet{Ilbert2009}. 
  The photo-z accuracy for the 24\,\micron\ sample is $\sigma=0.009$ at $i_\txn{AB} \le 22.5$ (with 5700 $z_\txn{spec}$) and 0.022 for the fainter sample (with 470 $z_\txn{spec}$).\footnote{The accuracy of the photometric redshifts is based on the normalized median absolute deviation \citep{Hoaglin1983}: $1.48\times Median(|z_s-z_p|/(1+z_s))$, where $z_s$ and $z_p$ are the spectroscopic and photometric redshifts, respectively.}
   
We limit our analysis to a maximum redshift $z\sim 1.3$ in order to derive reliable IR luminosities from the 24\,\micron\ flux density (see next Section). We reject AGN dominated sources according to their mid-IR properties based on the diagnostic of \citet{Stern2007}. 
 The selected sample consist of $\sim 16\,500$ star-forming galaxies and $\sim400$ evolved/passive galaxies, with  $S_{24}\ge 80 \, \mJy$ and $z \le 1.3$. The separation between star-forming and passive galaxies is based on the position of the galaxies in the rest-frame $(NUV-r)$ versus $(r-K)$ diagram as discussed in Section~\ref{sec:nuvrk} and Appendix~\ref{app:passive}. 
\subsubsection{A low-z  sample}  
 We complement the COSMOS field with a lower redshift sample selected from \citet{Johnson2007}, which  includes SWIRE observations  \citep{Lonsdale2003} in the Lockman Hole and the FLS regions. The multi-wavelength observations  combine the GALEX, SDSS and 2Mass photometry.  We restrict the sample to the sources detected in the three MIPS passbands  (24, 70 and 160\,\micron).
 % The physical parameters and luminosities are derived in the same way as our COSMOS sample, except for the IR luminosity, which we estimate by fitting the 8, 24, 70 and 160\,\micron\ bands  with a free scaling of the Far-IR SED (see next Section). 
 The final sample consists of $\sim730$  galaxies  with $z\le 0.2$  (mean redshift: $\overline{z} \sim0.11$) of which $\sim560$ are star-forming galaxies (with the same separation criterion as above). 
\subsection{Rest-frame quantities and  physical parameters}
\label{sec:phys_param}
\subsubsection{The infrared luminosity}
\label{sec:lir}
The total Infrared luminosity  ($\Lir$)  refers to the luminosity integrated from 8 to 1000\,\micron\  and is derived by using the code "Le Phare"\footnote{http://www.cfht.hawaii.edu/$\sim$arnouts/lephare.html} \citep{Arnouts1999, Ilbert2006}  combined with infrared SED templates of \citet{Dale2002}.
 For the low-z sample, the IR luminosity is estimated by fitting the 8, 24, 70 and 160\,\micron\ bands with a free scaling of the SEDs \citep[see][for details]{Goto2011}.
For the COSMOS sample,  the $\Lir$ is derived by extrapolating the 24\,\micron\ observed flux density  with  the star-forming  galaxy  templates of \citet{Dale2002}. 
While such SEDs are provided as a function of $60/100\,\micron$ luminosities ratio, we rescale the templates following the locally-observed dust temperature-luminosity relationship, so as to mimic a  luminosity dependence  similar to the SEDs of the libraries proposed by \citet{Chary2001} and \citet{Lagache2004}. In this scheme, the monochromatic luminosity at any given IR wavelength is linked to $\Lir$ by a monotonic relation, similar to the correlations that have been observed between the mid-IR emission and the total IR luminosity of galaxies in the local Universe \citep[e.g.,][]{Spinoglio1995, Chary2001, Takeuchi2003, Treyer2010, Goto2011}.
In this way, a 24\,\micron\ flux density observed at a given redshift is associated to a unique $\Lir$. Note that we did not consider any AGN SEDs in this work, since AGN-dominated sources were removed from our initial sample (see Sect.2.1.1). \\
Such extrapolations from mid-IR photometry have been widely used in the past, especially for the interpretation of the mid-IR deep field observations carried out with the {\it infrared Space Observatory} and the {\it Spitzer Space Telescope}. Given the large SED variations between the rest-frame emission probed at 24\,\micron\ and the peak of the IR spectral energy distribution where the bulk of the galaxy luminosity is produced, the estimate of $\Lir$ with this method yet depends on the assumed SED library.
 To quantify this effect, we use different IR star-forming galaxy templates of the literature, finding systematic differences of only $\la0.2$ dex up to $z \sim 1$ (e.g., see Fig.~7 of \citealt{Lefloch2005}).  Furthermore, stacking analysis of mid-IR selected sources with MIPS-70\micron\ and MIPS-160\micron\ data in COSMOS \citep{Lee2010} and other fields \citep[e.g.,,][]{Papovich2007,Bavouzet2008} has allowed accurate determinations of average IR luminosities for galaxies stacked by bin of 24\micron\ flux. These studies have shown very good agreement with the 24\micron\ extrapolated luminosities of star-forming galaxies up to $z \sim 1.5$, thus confirming the robustness of the technique.  Even more convincingly, direct far-IR measurements of the total IR luminosities for a 24\micron\ flux-limited sample of star-forming galaxies at $z \le 1.5$ with the {\it Herschel Space Observatory} have recently revealed a tight correlation with the luminosities extrapolated from 24\,\micron\ photometry and star-forming SED templates \citep{Elbaz2010}. The comparison between the two estimates  shows a dispersion less than 0.15 dex at $z \le 1$, demonstrating again the relevance of the method at least for mid-IR selected sources at low and intermediate redshifts, such as in our current study. For these reasons we restrict our analysis to redshifts lower than $z\sim 1.3$, where the method discussed above to derive the $\Lir$ is widely tested and robust.
%%%%%%%%%%%%%%%%%%%%%%%%%%%%%%
\begin{figure}
	\centering
\includegraphics[width=\hsize]{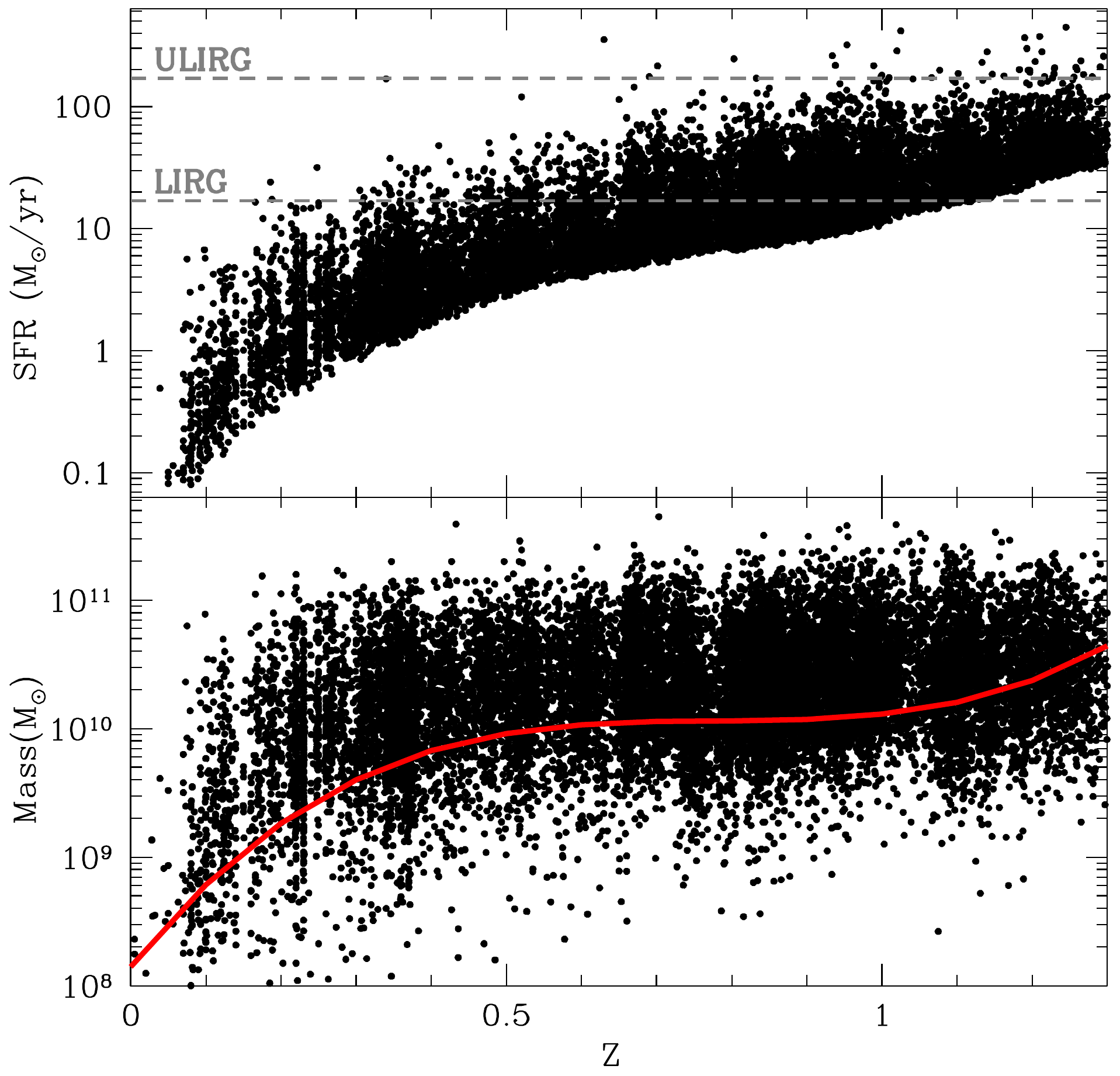}   
\caption[]{ Star Formation Rate (top panel) and  stellar mass distribution (bottom panel)
as function of redshift  for the  24\,\micron-selected  star-forming population. The horizontal gray lines in the top panel indicate the SFR thresholds corresponding to luminous and ultra-luminous infrared Galaxies (i.e., LIRGs and ULIRGs). The solid red line in the bottom panel indicates the 50 \% completeness limit [i.e., $\txn{M}_{50\%}(z)$ , see Section~\ref{sec:physic_param}]. }
\label{fig:mass}
\end{figure}
%%%%%%%%%%%%%%%%%%%%%%%%%%%%%%
%
\subsubsection{The other physical parameters and luminosities} 
\label{sec:physic_param} 
 We use the code "Le Phare"  combined with the population synthesis code of  \citet[, hereafter BC03]{BC03}  to derive the physical parameters for each galaxy. Details regarding the SED fitting are given in Appendix~\ref{app:sed}. We perform a maximum likelihood analysis, assuming independent Gaussian distributed errors, including all the available photometric bands from 0.15 to 4.5\,\micron . The physical parameters are derived by considering the median value of the likelihood marginalized over each parameter, while errors correspond to the 68 \% credible region.

We adopt the same approach as \citet{Ilbert2006} to derive the rest-frame luminosities (or absolute magnitudes). We use the photometry in the nearest rest-frame broadband filter to minimize the dependency to the k-correction. These luminosities are consistent within 10 \% with those derived according to the best fit template (smallest $\chi^2$)  or those provided by \citet{Ilbert2009}  based on a smaller set of empirical SEDs. Our results are not significantly affected by the adopted choices of luminosities.\footnote{We remind the readers that  systematic shifts in photometric passbands can propagate into the absolute magnitudes and then in the Equation 2 and 3 of this work. A change in calibration of the IRAC 3.6$\mu$m of $\sim$0.1 mag  will affect  $K_{ABS}$  by the same amount  and  the SFR by $\sim$0.06 dex.}

We note also that the great accuracy of the COSMOS photometric redshifts in our redshift domain allows us to neglect the impact of the photo-z errors in the quantities discussed here. However, we adopt spectroscopic redshifts when available. \\
 Throughout  this paper, the stellar masses always refer to the estimates from the SED fitting technique, assuming a Chabrier IMF truncated at 0.1 and 100 \Msun ,  while the total SFR is defined  as the sum of the unobscured ultraviolet and total IR luminosities. For the latter, we adopt a relation similar to that proposed by \citet{Bell2005} and adjusted for  a \citet{Chabrier2003} IMF:
\begin{equation}
\label{eq:sfr}
SFR^\txn{tot}[\MsunYr]= 8.6\ 10^{-11} \times (\Lir+2.3\times {\cal L}_{NUV}) ,  
\end{equation}
where the total IR luminosity  is defined as 
 $\Lir / L_\odot \equiv \int_8^{1000}d\lambda L(\lambda)$ and  ${\cal L}_{NUV}$ is the monochromatic NUV luminosity: ${\cal L}_{NUV} / L_\odot =\nu L_{\nu}(2300\,\AA)$.
 We adopt a factor of $\sim$2.3 instead of 1.9 as proposed by \citet{Bell2005} who uses the FUV luminosity.  According to stellar population models, our factor is more appropriated to correct the $L(2300\AA)$ in total UV luminosity.
The use of  the FUV luminosity  better traces the emission of short-lived, massive stars,  while in this work we use the NUV luminosity.
 The reason is that we can obtain a more accurate rest-frame NUV than FUV luminosity, thanks to the GALEX NUV and CFHT $u$-band observations at different redshifts. As shown by \citet{Hao2011}, this choice does not impact the reliability of the SFR estimates.  

 It is worth noting that, as shown in Appendix~\ref{app:sed}, the instantaneous SFR derived from the SED fitting is consistent within less than a factor of two with the SFR estimated with Equation~\eqref{eq:sfr}. Considering that the SED fitting relies only on the $0.15-4.5$\,\micron\ bands, this agreement between the different SFR estimates over at least 2 orders of magnitudes is remarkable.

Fig.~\ref{fig:mass} shows the SFR (top panel; as estimated with Equation~\eqref{eq:sfr})  and stellar mass (bottom panel) as a function of redshift for the 24\,\micron\ star-forming galaxies. As already shown by \citet{Lefloch2005}, below $z\le 0.5$ the population is composed of moderately star-forming galaxies with $\SFR \le 10 \,\MsunYr$. The fraction of luminous IR galaxies (LIRGs) gradually increases from $z=0.5$ to $z=1$ and becomes dominant at $z \ga 1$ in our sample. At all redshift, the fraction of ultra luminous IR galaxies (ULIRGs) is negligible. 

     Our sample is dominated by galaxies with $\log (\txn{M}/\Msun) \ga 9.5$.  To characterize how representative the 24\,\micron\ population is with respect to the entire star-forming (hereafter SF) population at a given mass and redshift, we define a 50 \% completeness  mass limit ($\txn{M}_{50\%}(z)$) as the stellar mass where the ratio  $\Phi_{SF}^{24\mu}(\txn{M},z)d\txn{M}\ /\  \Phi_{SF}^{All}(\txn{M},z)d\txn{M} \sim 0.5$; where $\Phi_{SF}^{24\mu}(\txn{M})$  and  $\Phi_{SF}^{All}(\txn{M})$ are the  $V_{max}$ weighted comoving volume densities of 24\,\micron- and K-selected  ($K\le23.5$) samples of star-forming galaxies respectively.
  Above this limit  ($\sim 10^{10} \ [10^{10.5}]\,\Msun$ at $z\sim1.1\  [1.3]$), we consider the  physical properties of the 24\,\micron\ population to be  representative of the whole SF sample. 
%
%%%%%%%%%%%%%%%%%%%%%%%%%%%%%%
\begin{figure}
	\centering
\includegraphics[width=\hsize]{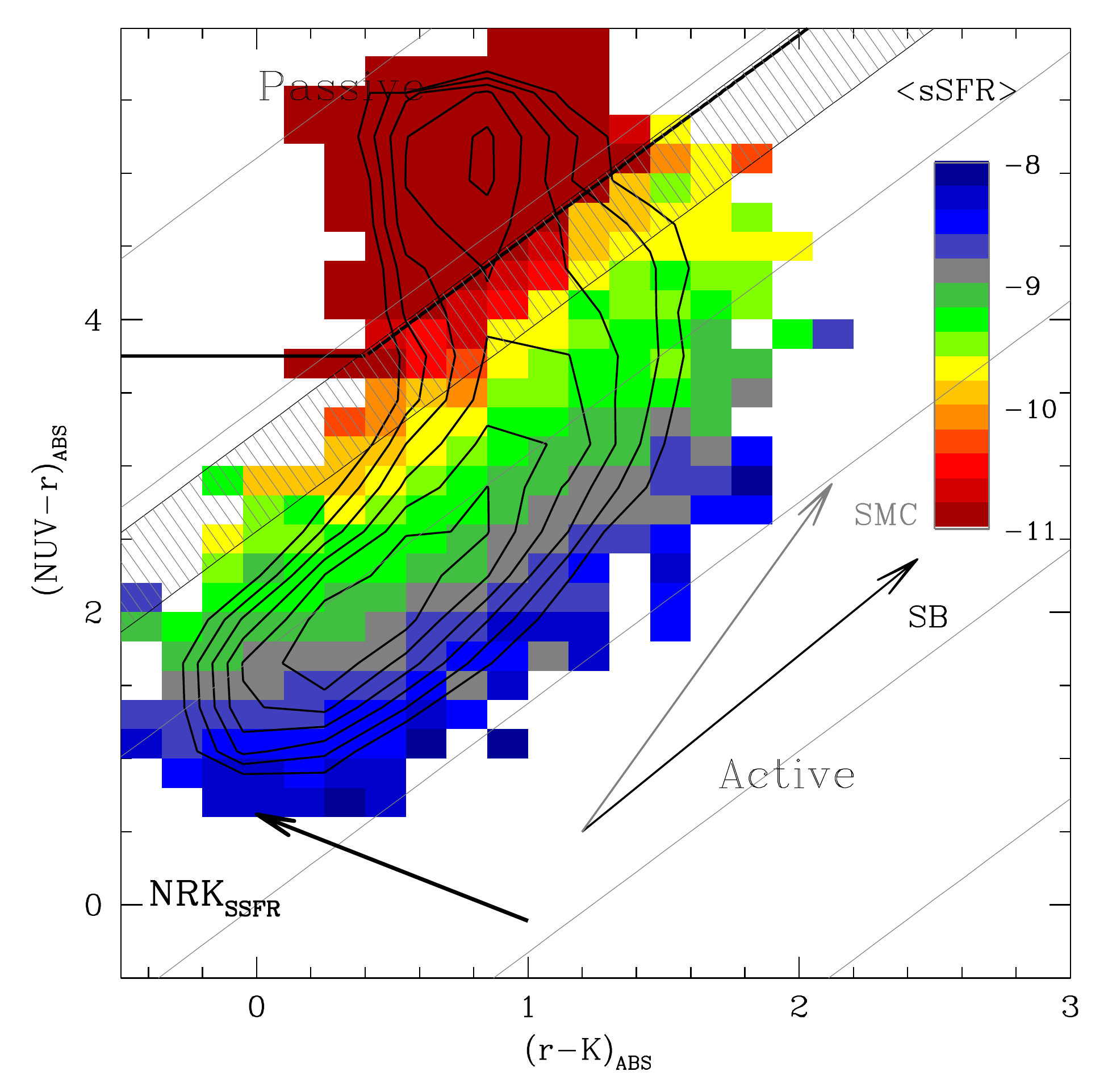}   
\caption[]{ Mean value of specific SFR (derived from the SED fitting,$\sSFRsed$ and color coded as indicated)  in the \NUVrK\  diagram for the entire COSMOS spectroscopic sample with $0.2\le z\le 1.3$. The density contours (thin black lines) are logarithmically spaced by 0.2 dex. The heavy black lines delineate the region of passive galaxies.  The shaded area defines the "intermediate" zone where
 the SFR estimates disagree between the different methods (UV+IR, NRK and SED fitting) as discussed in Section~\ref{sec:dep}. We show the attenuation vectors for starburst (SB) and SMC attenuation curves assuming  $E(B-V)=0.4$ and the vector $NRK_{ssfr}$ perpendicular to the starburst attenuation (see Section~\ref{sec:dep2}, {\it note the different dynamic ranges in x and y-axis warping the angles)}.
}
\label{fig:ssfr}
\end{figure}
%%%%%%%%%%%%%%%%%%%%%%%%%%%%%%%%%%%%%%%%%%
%     
\subsection{The $(NUV-r)$ versus $(r-K)$  color-color diagram} 
\label{sec:nuvrk}
 Recently, \citet{Williams2009} have shown that quiescent and star-forming galaxies occupy two distinct regions in  the  rest-frame $(U-V)$ versus $(V-J)$ color-color diagram  (i.e., $UVJ$)  and  validated their separation scheme with a morphological analysis \citep{Patel2011}. In the present work, we adopt the $(NUV-r)$ versus $(r-K)$ color-color (i.e., \NUVrK ) diagram to increase the wavelength leverage between current star formation activity and dust reddening. The \NUVrK\ diagram provides also an efficient way to separate passive and star-forming  galaxies.  In Fig.~\ref{fig:ssfr}, we show the mean sSFR derived from the SED fitting for the entire COSMOS spectroscopic sample with  $0.2\le z\le 1.3$. The density contour levels (black solid lines) reveal the presence in the top left part of the diagram of a  population with low specific SFR [$\log(\sSFRsed/\invYr) \la -10.5$], well separated from the rest of the sample. We define this region with the following criteria:   $(NUV-r)>3.75$  and  $(NUV-r)>1.37\times(r-K) + 3.2$. 
  In Appendix~\ref{app:passive} we discuss further the star-forming and passive galaxy separation based on the morphological information from HST imaging of the COSMOS field.    
%
%%%%%%%%%%%%%%%%%%%%%%%%%%%%%%
\begin{figure}
	\centering
\includegraphics[width=\hsize]{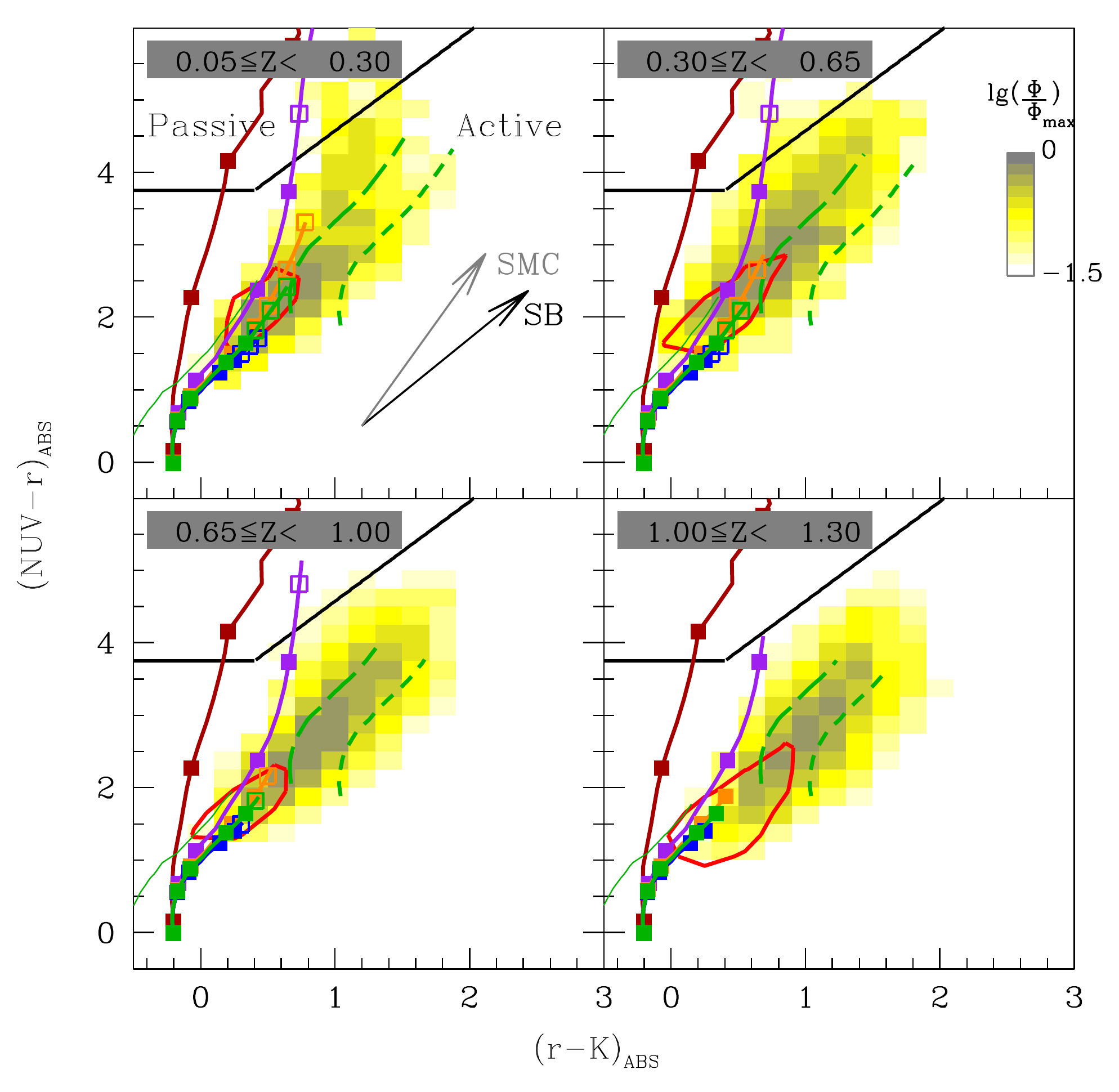}   
\caption[]{Observed density distribution of the 24\,\micron\ sample in \NUVrK\  diagram in four redshift bins, color coded (yellow-to-gray), in a logarithmic scale (step=0.15 dex)
 and normalized by the maximum density.  For comparison we show as a solid red line  the 50 \% contour density for the K-selected star-forming population ($K\le 23,5$). We overplot  tracks for BC03 models  with different e-folding times ($\tau=0.1$ Gyr, red line; 1 Gyr, purple; 3 Gyr, orange; 5 Gyr, green; 30 Gyr, blue). Symbols mark the model ages at  $t= 0.1,\ 0.5,\ 1,\  3,\  5$ Gyr (filled squares) and $t=6.5,\  9.0,\  12$ Gyr (open squares).  For the model with $\tau=5$ Gyr (thick green lines), we also show  the impact of subsolar  metallicity ($Z=0.2\,Z_{\odot}$: thin line) and dust attenuation  (SMC-like extinction law:  long dashed line and \citealt{Calzetti2000} law: short dashed line) assuming a reddening  excess $E(B-V)=0.4$.  The thick black line in the top-left of each panel delineates the region of quiescent galaxies.   }
\label{fig:nuvrk}
\end{figure}
%%%%%%%%%%%%%%%%%%%%%%%%%%%%%%%%%%%%%%%%%
%

 In Fig.~\ref{fig:nuvrk}, we  show the density distribution of the 24\micron\ sample in this \NUVrK\ diagram and the color tracks for five models with exponentially declining star formation histories.  The colors indicate models with different e-folding times $\tau= 0.1,\ 1,\ 3,\ 5,\ 30$ Gyr (red, purple, orange, green, blue lines respectively), while the filled and open squares mark different ages ($t= 0.1,\ 0.5,\ 1,\  3,\  5$ and $t=6.5,\ 9.0,\ 12$ Gyr, respectively).
 
The $(NUV-r)$ is a good tracer of the specific SFR, since the NUV band is sensitive to recent (i.e., $t \le 10^{8.5}$ yr) star formation and the $r$-band to old stellar populations \citep[e.g.,][]{Salim2005}.  Models with short star formation timescales  quickly move to the top side of the diagram, since star formation ceases early and the integrated light becomes dominated by old, red stars.  
 On the other hand, models with longer e-folding times show bluer colors at all ages, since they experience a more continuous star formation which replenishes the galaxy with young, hot stars emitting in the UV. This behavior has been widely used to separate the active and passive galaxy populations  based on   $(NUV-r)$ vs stellar mass (or luminosity) diagrams \citep[e.g.,][]{Martin2007, Salim2005}. 
 However, dust attenuation in star-forming galaxies can strongly alter the $(NUV-r)$ color, producing variations up to several magnitudes [e.g., $\Delta(NUV-R)\sim2$ for $E(B-V)\sim0.4$ and SMC law]. 
 Following  \citet{Williams2009}  we use a second color $(r-K)$, 
 %more sensitive to dust attenuation than to the age of the underlying stellar population, 
 which does not vary significantly with the underlying stellar population, even for passive galaxies [while $(NUV-r)$ does],   to partially break this degeneracy. 
  As  shown in Fig.~\ref{fig:nuvrk}, the models of spectral evolution span a much smaller range of  $(r-K)$ color than that observed in the $24\,\micron$ sample, unless dust attenuation is included.  We can qualitatively reproduce the observed ranges of $(NUV-r)$ and $(r-K)$ colors  by applying a reddening excess $E(B-V)\le 0.4-0.6$, assuming a starburst attenuation law or an SMC-like extinction curve.  
%
%    
%%%%%%%%%%%%%%%%%%%%%%%%%%%%%%
\begin{figure*}
%\begin{center}
%\centering 
\subfigure{ \includegraphics[width=0.45\hsize]{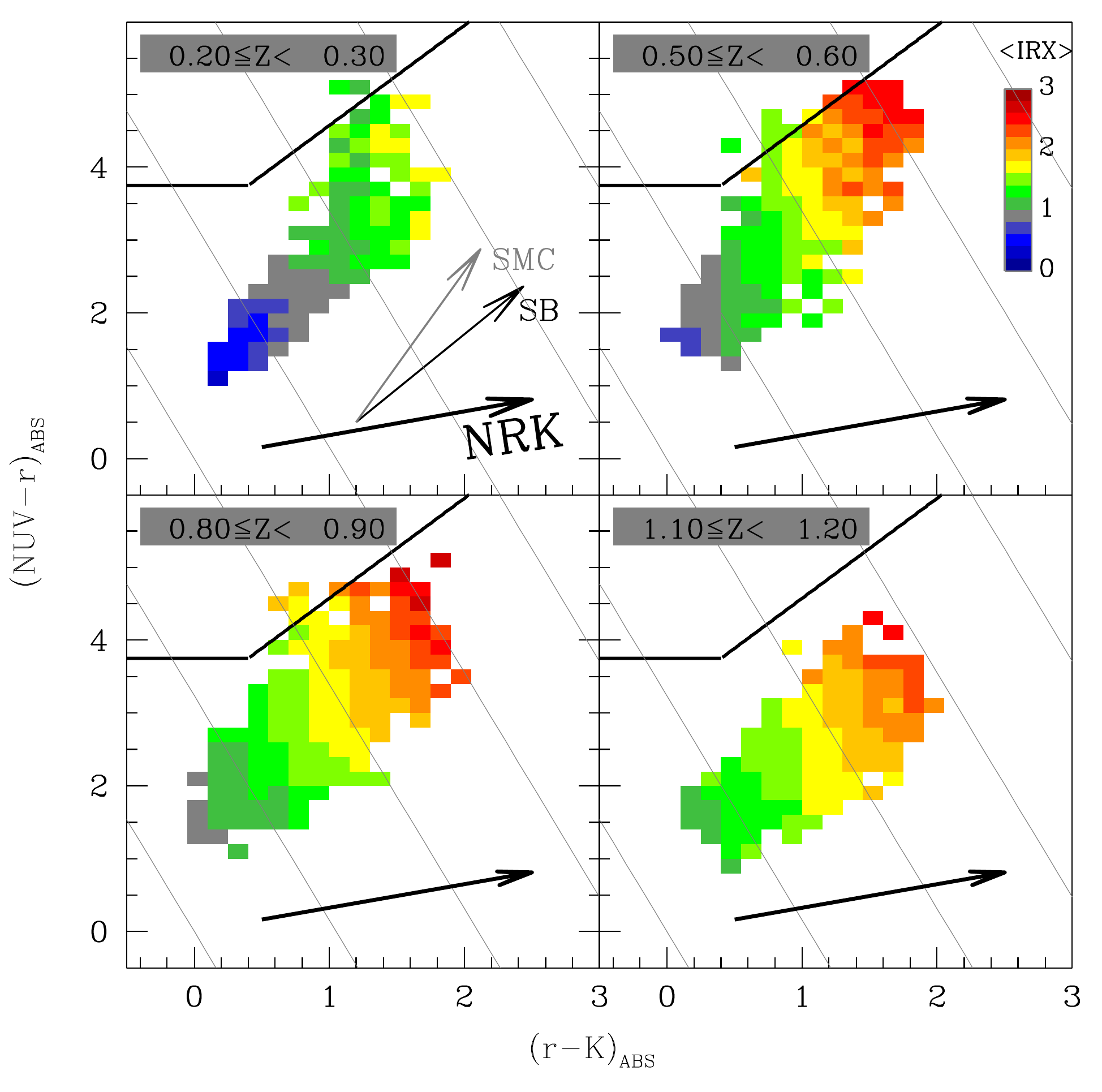} }
\subfigure{ \includegraphics[width=0.45\hsize]{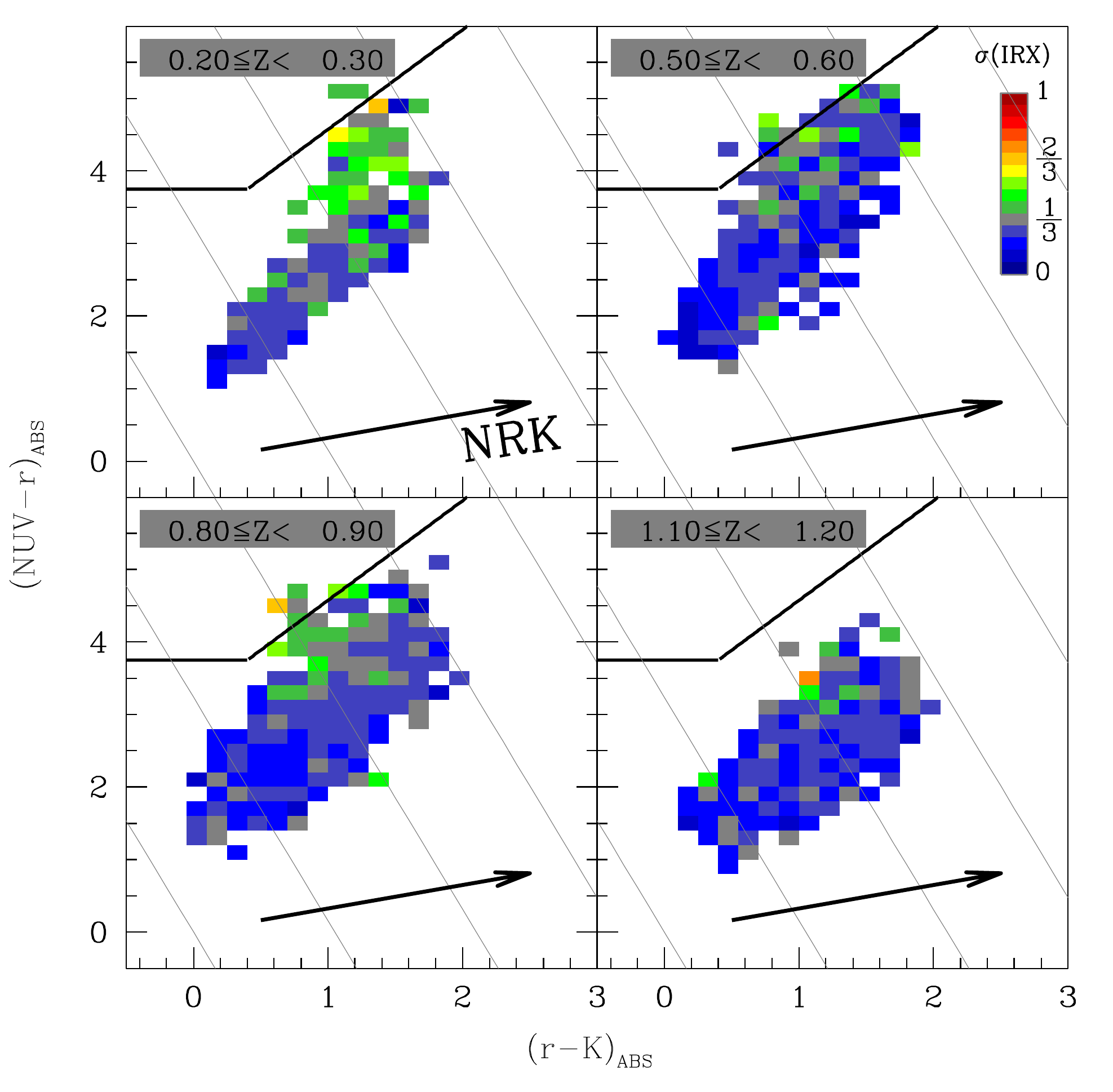}}
%%\end{center}
\caption[]{Infrared excess (IRX) in the $(NUV-r)$ versus $(r-K)$ diagram.  Left figure:  volume-weighted mean IRX (\irx) for  the 24\,\micron-selected sources in 4 redshift bins, color coded in a logarithmic scale (shown in the top right panel). We overplot in the top-left panel the attenuation vectors for starburst and SMC attenuation curves assuming  $E(B-V)=0.4$.
 In each panel, we overplot the vector \NRKv\  (black arrows) and its perpendicular lines (gray solid lines) corresponding  to \NRK\  in the range  0 to 4.
 ({\it note the different dynamic ranges in x and y-axis warping the angles)}.  
Right figure: dispersion around the mean ($\sigma(IRX)$), color coded in a logarithmic scale.}
\label{fig:irx}
\end{figure*}
%
%%%%%%%%%%%%%%%%%%%%%%%%%%%%%%
\section{Infrared excess of star-forming galaxies in the \NUVrK\ diagram}
\label{sec:infrared_exc}
\subsection{Definition of the IRX and \NRK\ parameters}
We now explore  the behavior of the infrared excess in the \NUVrK\ diagram. We define  the infrared excess as  $IRX=\Lir/{\cal L}_{NUV}$, where $\Lir$ and ${\cal L}_{NUV}$ are defined in Section~\ref{sec:physic_param}. 
  This  ratio  is weakly dependent on the age of the  stellar population,  dust geometry and nature of the extinction law  \citep{Witt2000}.  
 
Fig.~\ref{fig:irx} shows on the left,  the volume weighted mean IRX (\irx, color-coded in a logarithmic scale shown in  the top right panel), and, on the right,  the dispersion around the mean ($\sigma(IRX)$) in  the \NUVrK\  diagram in four redshift bins.  At any redshift,  \irx\ increases  by $\sim1.5-2$ dex from the blue side (bottom-left) to the red side  (top-right) of the diagram, while the dispersion around the mean remains  small and constant ($\le0.3$ dex). We note also the presence of stripes of constant \irx\ values, which we discuss in Section~\ref{sec:discu}. The presence of such stripes allows us to describe the variation of \irx\  in the \NUVrK\  diagram with a single vector perpendicular to those stripes.
 This vector, hereafter called \NRKv, can be defined as a linear combination of rest-frame colors $\NRKv(\phi)= \sin(\phi)\times(NUV-r) + \cos(\phi)\times (r-K)$, where $\phi$ is an adjustable parameter which we require to be perpendicular to the \irx\ stripes. 
 We empirically estimate $\phi$ in each redshift bin ($\Delta z=0.1$) by performing a linear least square fit: $IRX=f[NRK(\phi)]$ (the linear approximation is justified in the next Section). Since the dispersion $\sigma[IRX(\phi)]$ reaches a minimum when the vector $\NRKv(\phi)$ is perpendicular to the stripes, we minimize $\sigma[IRX(\phi)]$ to find the best-fit angle, $\phi_b$. We find $15^{\circ} \le \phi_b \le 25^{\circ}$ in all redshift bins, with a median and semi-quartile range $\phi_b=18^{\circ} \pm 4^{\circ}$. Given the narrow distribution of $\phi_b$ and the fact that a change up to $\Delta \phi\pm 7^{\circ}$ does not affect our results, we adopt a unique definition for the vector \NRKv\  at all redshifts by fixing $\phi_b=18^{\circ}$,
 \begin{equation}
 \label{eq:nrk}
  \NRKv=  0.31\times (NUV-r) + 0.95\times (r-K)
 \end{equation}
In Fig.~\ref{fig:irx} we show the vector \NRKv\  (black arrows) and a number of lines perpendicular to it (gray  lines) corresponding to constant value of \NRK\ in the range $0 \le \NRK \le 4$.  
We note that \NRKv\  has a different orientation in the \NUVrK\ diagram with respect to the starburst and SMC dust attenuation vectors, as we discuss in Section~\ref{sec:discu}. 
\subsection{The calibration: \irx\ versus \NRK} 
\begin{figure*}
	\centering
\includegraphics[width=\hsize]{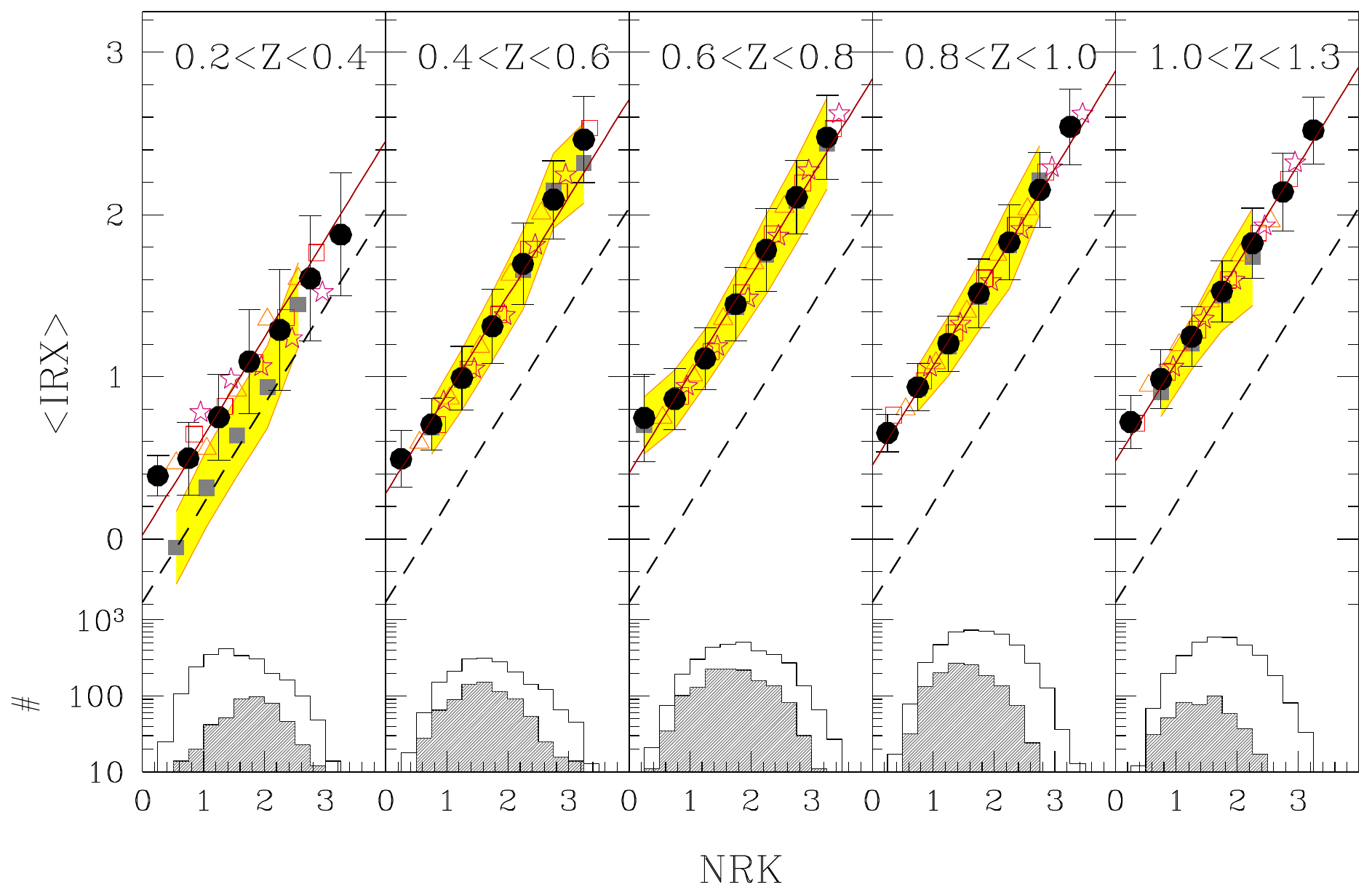}
\caption[]{Volume weighted mean IRX (\irx, on a logarithmic scale) as a function of \NRK\  in five redshift bins. The COSMOS sample based on photometric redshifts  is shown as solid black circles. Gray squares and yellow shaded area indicate the mean IRX and dispersion for the spectroscopic sample of local galaxies of \citet{Johnson2007} (left-most panel) and the COSMOS spectroscopic sample (four right-most panels). The solid red line indicates the predictions of Equation~\eqref{eq:irx} for \irx\ as a function of \NRK\  at the mean redshift of the bin, while the dashed line refers to $\overline{ z}\sim 0.11$ the mean redshift of the \citet{Johnson2007} sample.
 We overplot the different mass selected samples: $9.5\le \log(\txn{M}/\Msun)\le 10.0$ (open triangles) ;  $10.0\le \log(\txn{M}/\Msun)\le 10.5$ (open squares) ;  $10.5\le \log({M}/\Msun)\le 11.5$ (open stars). 
 The distributions of \NRK\  for the total (solid lines) and spectroscopic samples (shaded histograms) are shown in the bottom part of each panel.  
 }
\label{fig:irxnrk}
\end{figure*}

 With the definition of  \NRKv, we can now derive the relationship between  \irx\   and  \NRK. In Fig.~\ref{fig:irxnrk} we measure the mean IRX values (\irx)  and the associated dispersions  per bin of \NRK\ for the 24\,\micron\ sample (solid black circles) in different redshift bins.  At high redshift, a tight correlation is observed with a small dispersion ($\sigma\le 0.3$) compared to the evolution of \irx\ ($\Delta\irx\ \sim 2$ dex). At $z\ge 0.4$, we obtain almost the same results for the spectroscopic sample (gray squares with yellow shaded region). Although this sample is 1/10 of the 24\,\micron\ sample, we observe the same dispersions,  suggesting that we are  dominated by an intrinsic, physical scatter rather than Poisson noise. \\
  In the first panel ($0.2\le z\le 0.4$),  we  also include the lower redshift sample from \citet{Johnson2007} ($\overline{z}\sim 0.11$, gray squares with yellow shaded region).
 While the \irx\ vs \NRK\ relation remains the same as at higher redshift, the dispersion increases in both of the low-z samples. 
 This is related to the increasing fraction of less active galaxies with a higher contribution of evolved stars  affecting the  IR luminosity  and contributing to blur the correlation between IRX and NRK \citep[e.g.,][]{Cortese2008}.\\
Finally we measure the \irx\ vs \NRK\ relation for three stellar mass bins. We observe  no significant difference in the relation derived in each bin and the global one. This, along with the SED reconstruction analysis shown in Appendix~\ref{sec:sed},  supports our assumption of neglecting the dependence on stellar mass in the calibration of \irx\ vs \NRK\  in the mass  range considered here. 
%
%%%%%%%%%%%%%%%%%%%%%%%%%%%%%%
%%%%%%%%%%%%%%%%%%%%%%%%%%%%%%
%

In the parametrization of  \irx\  as a function of \NRK\ and redshift, we assume that  the  two quantities can be separated:
\begin{equation}
\label{eq:irx}
\log[\irx(z,\NRK)]= f(z) + a_{N}\times  \NRK 
\end{equation} 
where $f(z)$ is a third-order polynomial function describing the redshift evolution, and $a_N$ a constant which describes the evolution with \NRK.
%
%    
%%%%%%%%%%%%%%%%%%%%%%%%%%%%%%
\begin{figure}
%	\centering
\includegraphics[width=\hsize]{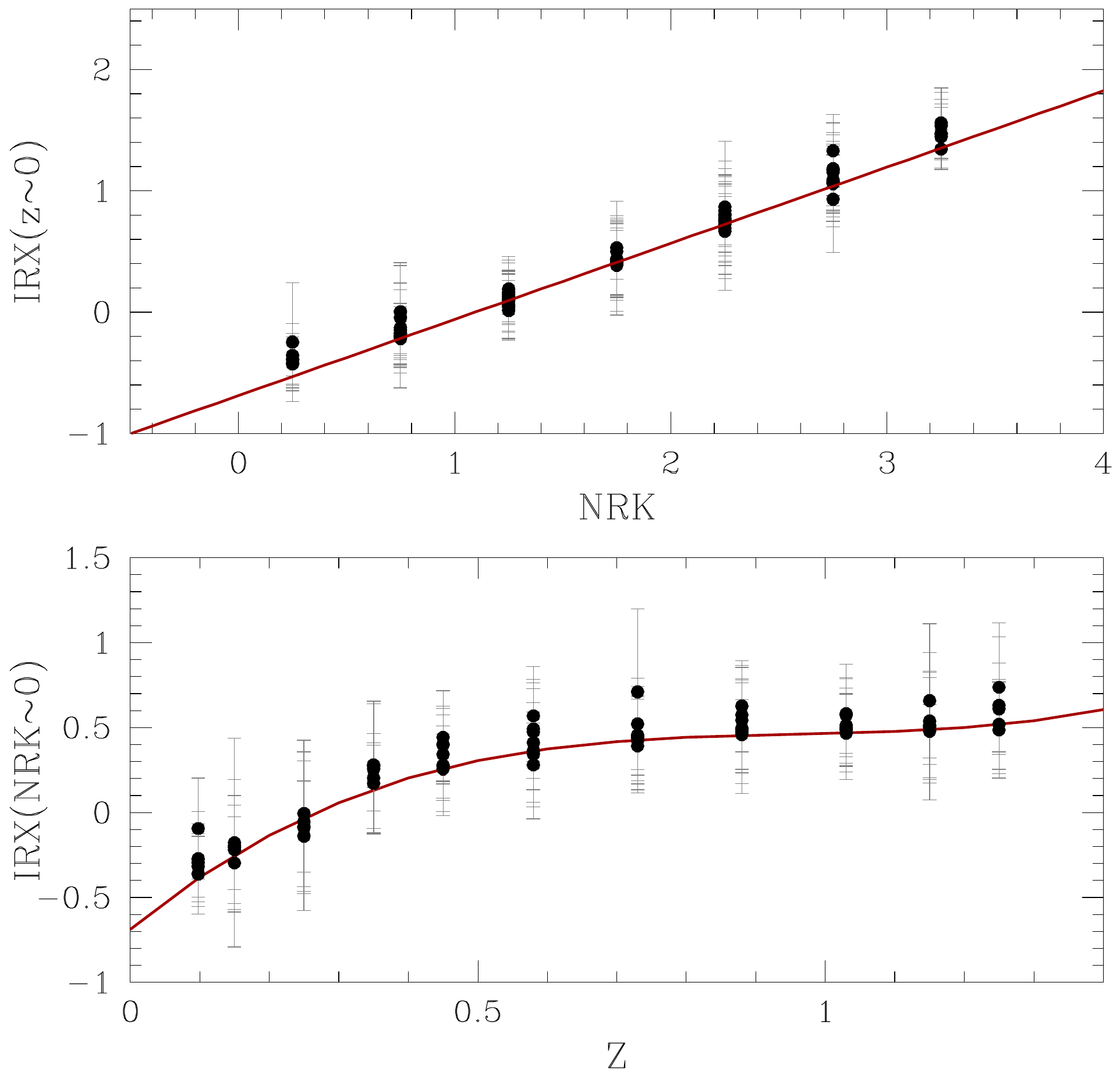}
\caption[]{Illustration of the analytical parametrization of the relation $\log[\irx]=f(\NRK, z)$.
 Top panel: linear fit as a function of \NRK\  after rescaling \irx\ at $z=0$. Bottom panel: polynomial fit as a function of redshift, after rescaling \irx\  at \NRK=0. }
\label{fig:calib}
\end{figure}
%%%%%%%%%%%%%%%%%%%%%%%%%%%%%%%%%%%%%%%%%%
%
 We bin the data in redshift ($\Delta z=0.1$)  and \NRK\ ($\Delta NRK=0.5$) and perform a linear least square fit to derive the five free parameters. We obtain for the redshift evolution $f(z)=a_0 +a_1 . z +a_2 . z^2+ a_3 . z^3$ with $a_0=-0.69\pm0.06$\,; $a_1=3.43\pm0.33$\,; $a_2=-3.49\pm0.55$\,; $a_3=1.22\pm0.28$, and for the dependence on NRK $a_N=0.63\pm0.01$.\\
 The resulting fits from this calibration are shown in Fig.~\ref{fig:irxnrk} (solid red lines). We show, in the top panel of  Fig.~\ref{fig:calib},  the linear fit of  \irx\ as a function of \NRK\ after rescaling all the \irx\ values at  $z=0$ and in the bottom panel,  the polynomial fit  of \irx\ vs $z$ after rescaling all the \irx\ values at  $\NRK=0$ 
 The small uncertainty in the slope parameter $a_N$ supports our initial choice of a linear function to describe the relation between \irx\ and \NRK.   We note that due to adoption of a polynomial function, the redshift evolution is only valid in the range well constrained by the data ($0.1\le z\le 1.3$) and should not be extrapolated outside this range. \\
 A simple interpretation of the redshift evolution of \irx, at fixed \NRK, is the aging of the stellar populations. In fact, Fig.~\ref{fig:nuvrk} shows that, at a fixed position in the \NUVrK\ diagram (or NRK value), a galaxy at higher redshift, which has a younger stellar population, needs a larger dust reddening than a galaxy at lower redshift, which hosts older, intrinsically redder stars. This effect being more pronounced between $0\le z\le 0.5$ where the universe ages by $\Delta T\sim 5$ Gyr, compared to 4 Gyr in between $0.5\le z\le 1.3$, and which is also enhanced by the global decline of the SF activity in galaxies with cosmic time.
%
%%%%%%%%%%%%%%%%%%%%%%%%%%%%%%%%%%
% 
%%%%%%%%%%%%%%%%%%%%%%%%%%%%%%%
\section{The infrared luminosity and SFR estimated from \NRK\  vector}
%
%%%%%%%%%%%%%%%%%%%%%%%%%%%%%%
\begin{figure*}
	\centering
\subfigure{ \includegraphics[width=0.3\hsize]{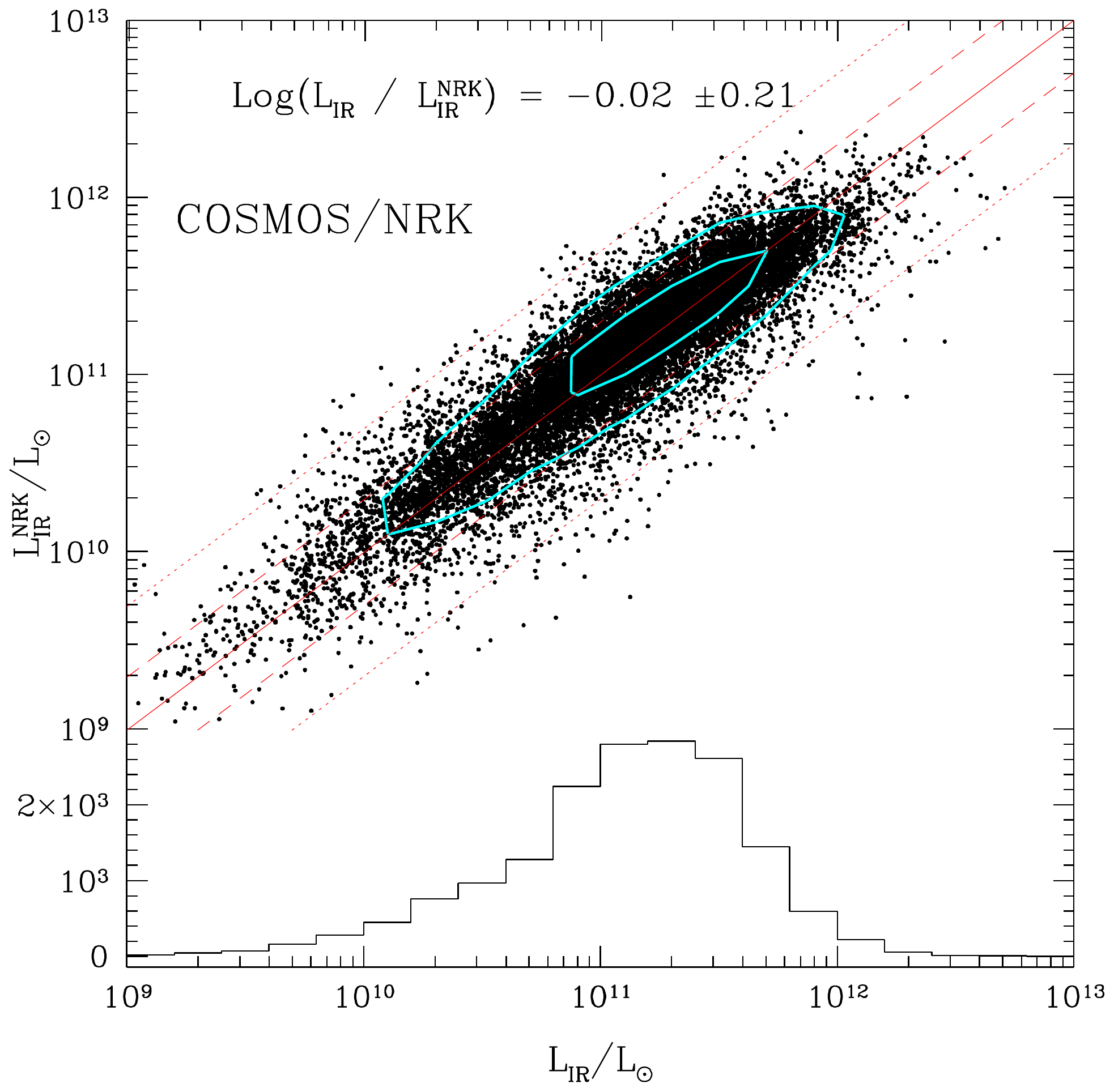} }
\subfigure{ \includegraphics[width=0.3\hsize]{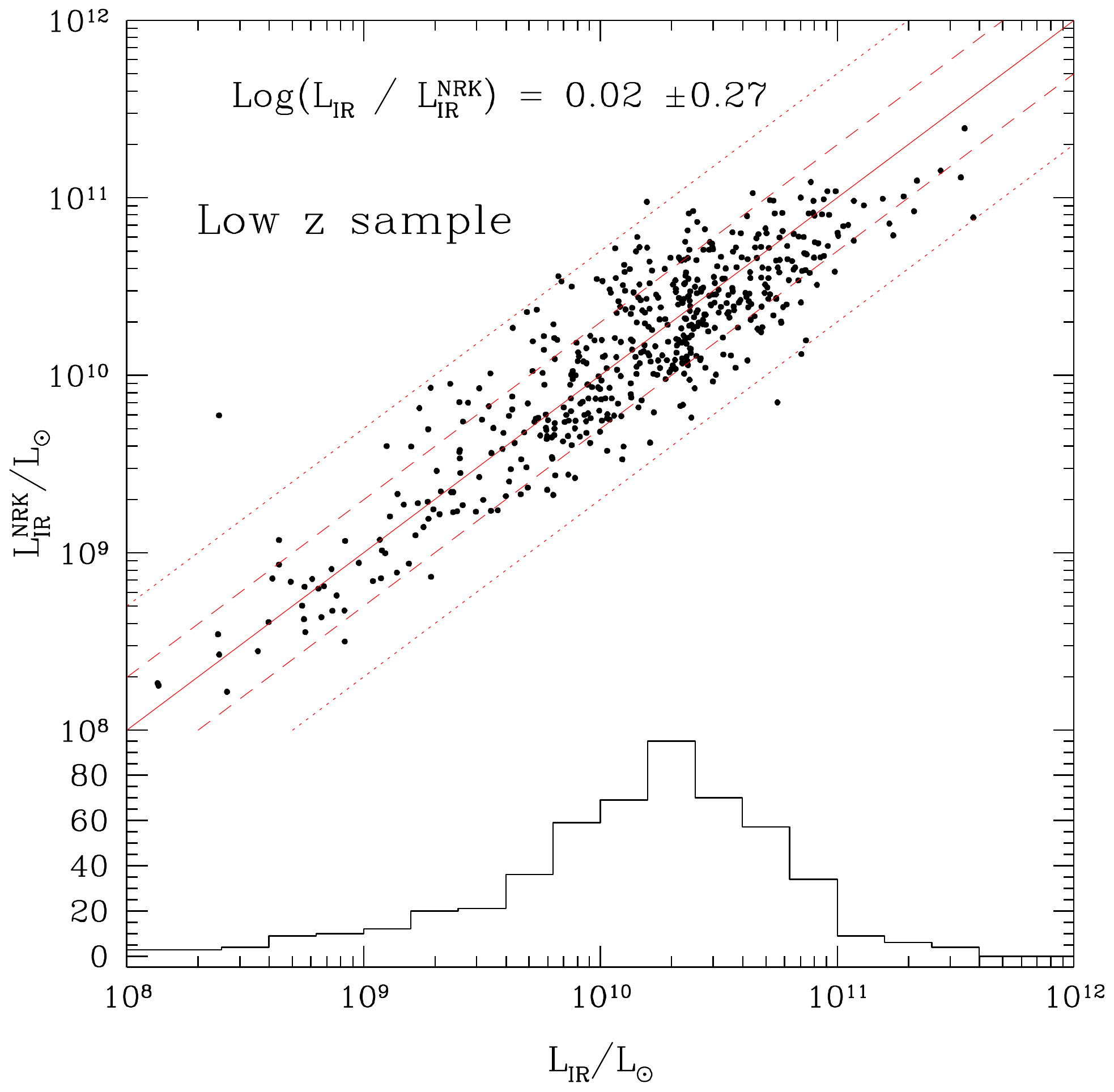} }
\subfigure{ \includegraphics[width=0.3\hsize]{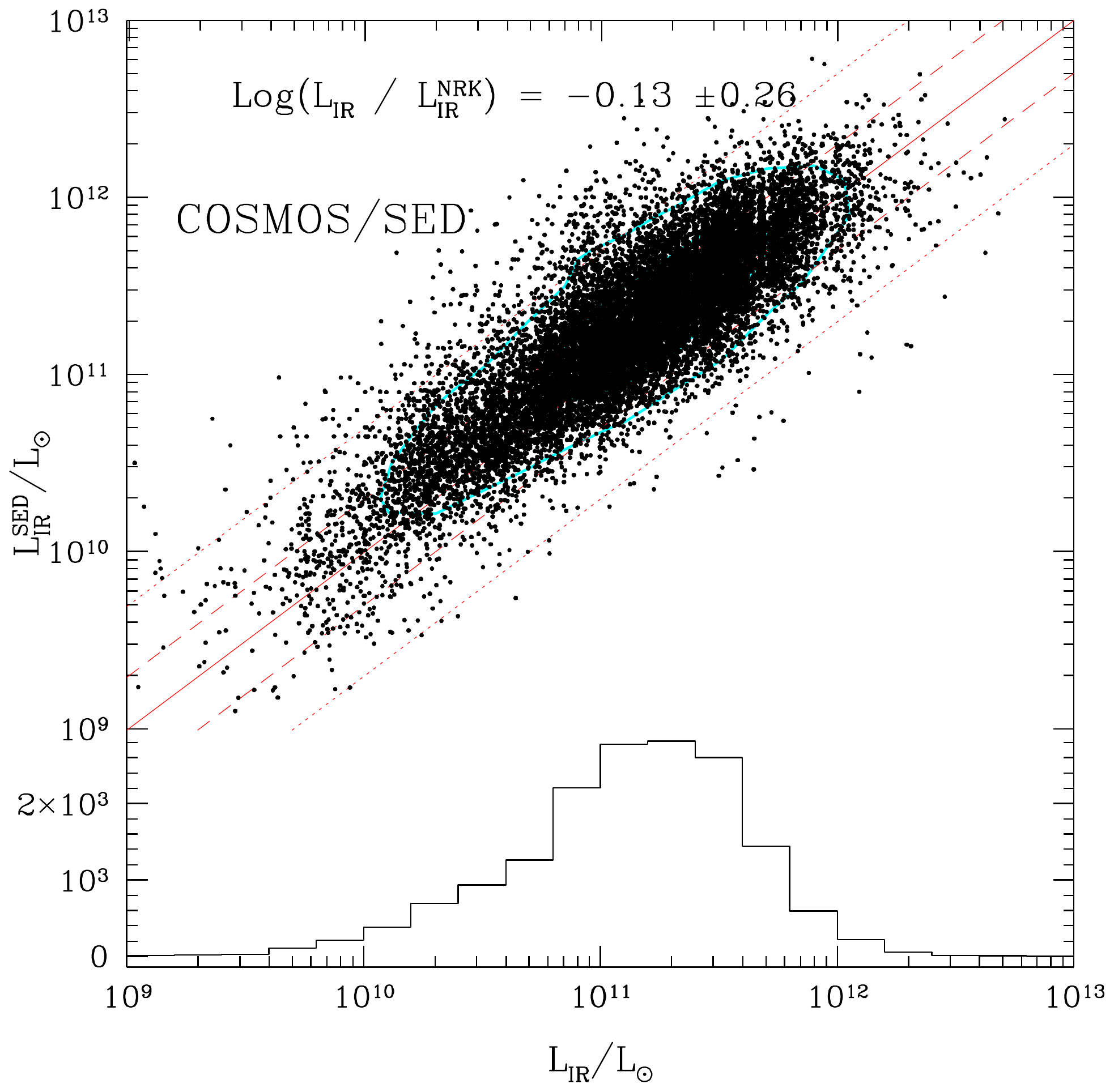} }
\caption[]{ Comparison of the infrared luminosity $\Lir$ estimated from the 24\,\micron\ luminosity (COSMOS sample) or 8, 24, 70 and 160\,\micron\ luminosities (local  sample) with that estimated with \NRK\ ($\LirNRK$), for the COSMOS (left panel) and SWIRE (middle panel) samples. We also compare with the one estimated from the SED fitting ($\LirSED$, right panel). In each panel, the dashed and dotted lines refer to a factor ratio of 2 and 5 respectively. The distributions of $\Lir$  are shown in the bottom part of each panel. The numbers in each panel refer to the logarithm of the mean and dispersion of the luminosity ratio for the star-forming galaxies. 
%The open triangles show galaxies with true sSFR lower than the  sSFR limit threshold where the  NRK relation over-estimates the sSFR by a factor 2. 
 }
\label{fig:dirx0}
\end{figure*}
%%%%%%%%%%%%%%%%%%%%%%%%%%%%%%%%%%%%%%%%%%
%
The relationship  between IRX  and  the vector \NRKv allows us  to predict the IR luminosity for each galaxy according to its  $NUV,\ r,\ K$ luminosities and redshift  as follows $\LirNRK = {\cal L}_{NUV} \times \irx(z,\NRK) $, where  \NRK\  and \irx\ are estimated from Equation~\eqref{eq:nrk} and \eqref{eq:irx}, respectively. 
\subsection{Comparison with the reference $\Lir$} 
In Fig.~\ref{fig:dirx0} (left and middle panels) we compare the  IR luminosity predicted with the \NRK\ method   ($\LirNRK$) with our reference IR luminosity derived from the mid/Far-IR bands ($\Lir$, see Section~\ref{sec:lir}).
 The IR luminosities estimated with the two methods for the $\sim16\,500$ star-forming galaxies in the  COSMOS 24\,\micron\  sample (left panel) agrees with almost no bias and a dispersion of $\sim0.2$ dex over the entire luminosity range. Less than 1 \%  of the galaxies shows a difference larger than a factor of 3. The prediction of the IR luminosity for the LIRG population, which dominates the COSMOS sample (the 24\,\micron\ population peaks at  $\Lir\sim 2\ 10^{11}L_{\odot}$), is excellent,  considering the small number of parameters involved in the \NRK\ method.\\
 At low redshift, the prediction of the IR luminosity for the SWIRE galaxies (middle panel), which are  10 times less luminous than the 24\,\micron\ COSMOS population  ($\Lir\sim 2 \ 10^{10} L_{\odot}$), shows a larger scatter ($\sigma \sim 0.27$ dex). This reflects the larger dispersion observed in Fig.~\ref{fig:irxnrk}  which is likely related to the wider range of galaxy properties at low z. %  as  well as the poor constraint of the polynomial fits at  very low redshift ($z<0.1$).
 \cite{Johnson07a} have also modeled the IRX as a function of  different rest-frame colors and $D_n(4000)$ break for the SWIRE galaxies (see their Table 2 and Eq. 2). The fit residuals for their predictions of IRX vary from $\sigma(IRX)\sim$ 0.27 to 0.36,  depending  on the adopted colors.  Even with the  use of the  $D_n(4000)$ break as a dust-free indicator of star formation history, they do not achieve a better accuracy than that obtained with the  method presented in this work. \\
 Finally, our method provides a smaller dispersion in the prediction of the IR luminosity  than the dust luminosity obtained from the SED fitting  (Fig.~\ref{fig:dirx0} right panel), where we obtain a dispersion of $\sim$ 0.26 dex and a bias of $\Delta\sim -0.13$. 
 We note that the $\Lir$ derived from SED fitting is  computed by integrating all the stellar photons absorbed by dust according to the adopted attenuation law and reddening excess (see Appendix~\ref{app:sed}), while the reference $\Lir$ is based on the extrapolation of  the 24\,\micron\ luminosity.  Both measurements suffer from independent source of uncertainties, thus the bias in the IR luminosity may be related to either methods. However, the small scatter (less than a factor 2) and  bias show that the SED fitting is a robust and reliable method to estimate the total IR luminosity, when accurate observations on a wide enough wavelength are available \citep[e.g., ][]{Salim2009}. This is indeed the case for the COSMOS dataset used here (i.e., 31 passbands from Far-UV to Mid-IR: $0.15 \le \lambda\le 4.5\,\micron$).
\subsection{Dependence of the predicted IR luminosity with physical galaxy properties}
\label{sec:dep}
\begin{figure}
	\centering
\includegraphics[width=\hsize]{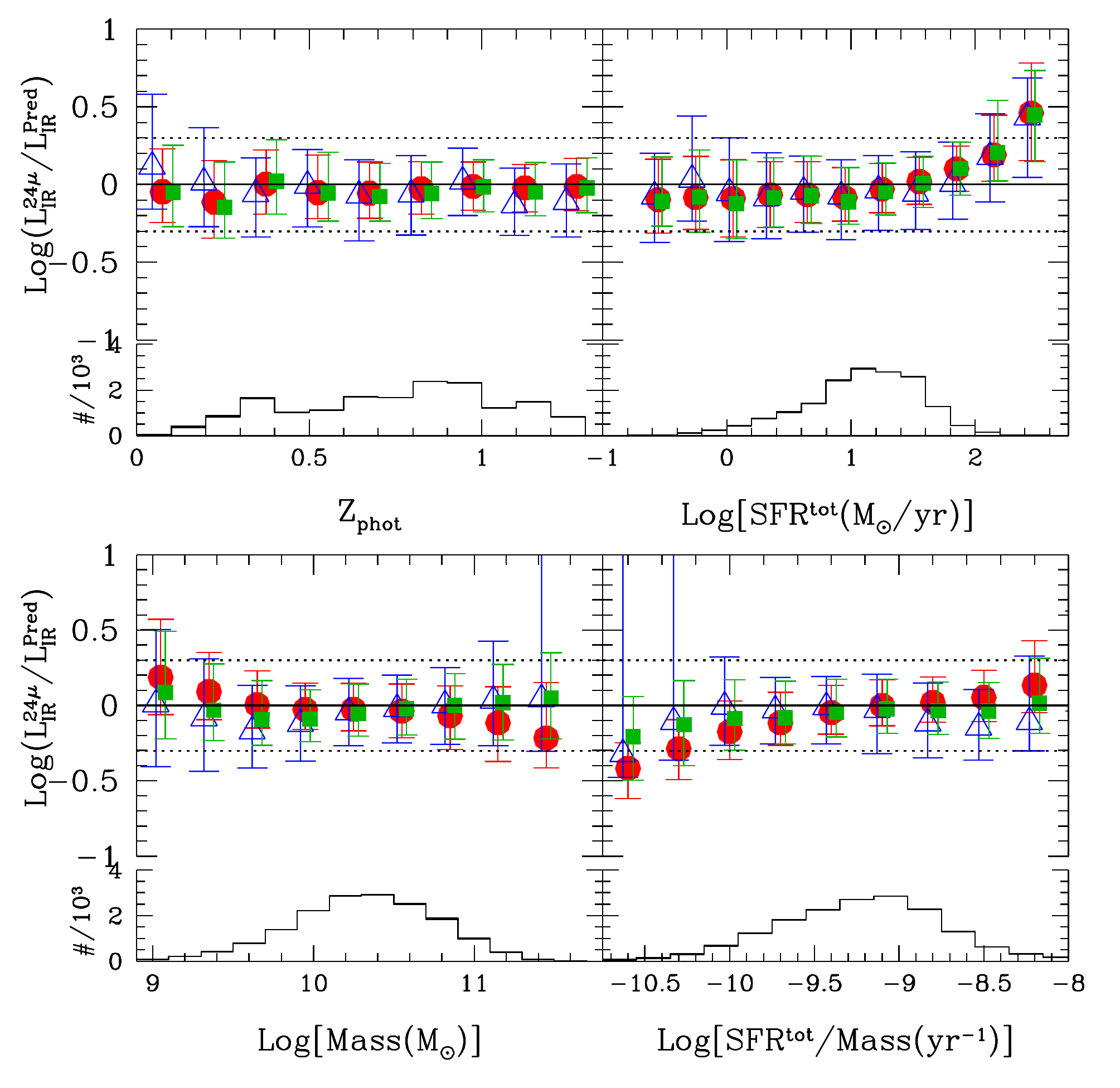}
\caption[]{Median and dispersion of the difference (i.e., residuals) between the IR luminosity based on the 24\,\micron\ flux (i.e., $L_\txn{IR}^{24\mu}$) and that predicted with different methods, as a function of different galaxy physical parameters. In each panel red-filled circles refer to the \NRK\ method, green squares to the \NRK\ sSFR-corrected and open-blue triangles to the SED fitting (open-blue triangles). Top-left panel shows the residuals as  function of redshift, top-right of total SFR, bottom-left of stellar mass and bottom-right of specific SFR. The number distribution for each physical parameter is shown as an histogram in the lower part of each panel.
}
\label{fig:dirx1}
\end{figure}

Despite the use of a single vector, \NRKv,  and the absence of mass dependency, our recipe provides a reasonable estimate of the total IR luminosity $\Lir$.  However, galaxies with physical properties that deviate from the bulk of the 24\,\micron\ population may be less accurately described in this framework.  To test this issue and determine the range of validity of the method, we show in Fig.~\ref{fig:dirx1} the residuals (i.e., the difference) between the reference (i.e., $\LirMU$) and predicted IR luminosity as a function of redshift (top-left panel), total SFR (top-right), stellar mass (bottom-left) and specific SFR (bottom-right). The total SFR refers to the (UV+IR) SFR (see Equation~\ref{eq:sfr}), while the stellar mass is obtained with the SED fitting (see Section~\ref{sec:physic_param}). 
 We perform this comparison for the $\Lir$ predicted with both the \NRK\ and SED fitting methods.  

 Top-left panel of Fig.~\ref{fig:dirx1} does not reveal any bias with redshift in the IR luminosity predicted with both methods. This is expected by construction for the \NRK\ method, while it confirms the good performance of the SED fitting technique despite  the larger scatter than the one obtained with the \NRK\ method.

Top-right panel of Fig.~\ref{fig:dirx1} shows that the residuals as a function of the total SFR are relatively stable, with an almost zero bias and a dispersion of $\sim$0.2 dex for $\SFR\le 100\,\MsunYr$. At higher SFR, where galaxies approach the ULIRG regime, the residuals start to deviate from zero, indicating that both the \NRK\ and SED fitting methods under-predict the reference IR luminosity.
The difference with the \NRK\ predictions is likely to be caused by the rarity of these highly star-forming objects in our redshift range \citep[$0 \le z \le 1.3$, see also][]{Lefloch2005}, which makes these objects under-represented in the \NRK\ calibration.
The disagreement with the prediction of the SED fitting may have a different origin. As shown in Appendix~\ref{app:sed}, the starburst attenuation law is favored at high SFR, however ULIRGs are found to deviate from Meurer's relation \citep{Reddy2009, Reddy2012a} and exhibit a higher IRX at a given UV-slope. This effect has been observed at higher redshift ($z\ge 1.5$), but it could also be the reason of the under-estimate of $\Lir$ by the SED fitting technique at lower redshift.  We also note that our e-folding SF history models used in the SED fitting may be too simplistic and models including burst episodes would be more appropriated for these actively star-forming galaxies. 
 Finally, we also note that the reference $\Lir$, based on the extrapolation of the 24\,\micron\ luminosity, may reach its limit of validity in this regime. Indeed, the conversion of the 24\,\micron\ luminosity to total IR luminosity becomes uncertain for ULIRGs (i.e., for $\Lir\ge 10^{12}L_{\odot}$, see \citealt{Bavouzet2008, Goto2011}). Also, the merger nature of ULIRGs at low z \citep{Kartaltepe2010} makes predictions inaccurate in absence of a complete set of far-IR observations.

Bottom-left panel of Fig.~\ref{fig:dirx1} do not reveal any bias with stellar mass for the IR luminosity predicted with the SED fitting technique, while the dispersion increases at the extreme sides. 
On the other hand, the \NRK\ method under-estimates $\Lir$ for stellar mass $\log(\txn{M}/\Msun)\la 9.3$. As shown in \citet{Lefloch2013} from the analysis of a mass-complete sample obtained with stacking techniques of Herschel-SPIRE data at 250/350/500\,\micron, the NRK calibration presented in this work needs to be modified for galaxies with $\txn{M} \la 10^{9.5}\,\Msun$.
At high stellar mass [i.e., $\log(\txn{M}/\Msun)\ga 11$] the \NRK\ method over-estimates $\Lir$. This reflects the correlation between stellar mass and sSFR, where quiescent galaxies become the dominant population at high stellar masses (see below).

Bottom-right panel of Fig.~\ref{fig:dirx1} shows that the \NRK\ method systematically (as reflected by the small scatter) over-estimates $\Lir$ at low sSFR  [i.e., $\log(\sSFR/\invYr)\le-10$].  We investigate this bias in more details in Appendix~\ref{app:dep_nrk} and find that the sSFR derived with NRK saturates at such low sSFR, while the reference sSFR keeps decreasing. We therefore propose a simple analytical correction for this bias in order to reconcile the two sSFR estimates, which can then be used to correct $\LirNRK$.  The results for this $\LirNRK$ sSFR-corrected are shown in Fig.~\ref{fig:dirx1} as filled green squares. 
The correction provides a better match to the 24\,\micron\ parent distribution across the entire range of stellar mass and sSFR, but at the cost of a larger dispersion for the global sample (i.e., from $\sigma=0.21$ to $\sigma=0.24$). It is worth noting that the reference sSFR for quiescent galaxies might also suffer from systematic error. In fact, recent analysis with Herschel data have reported evidences for a warmer dust temperature in early- than in late-type galaxies \citep{Skibba2011, Smith2012}. This will result in an over-estimation of $\Lir$ as derived from the 24\,\micron\ luminosity (by adopting a too high $\Lir/L_{24\mu}$ ratio). Combined with the UV luminosity partially produced by evolved stars, these two effects can lead to an over-estimate of our reference SFR (based on UV+IR). As seen in Fig.~\ref{fig:dirx1}, the $\Lir$ comparison with the SED fitting method shows a large dispersion at low sSFR. The reason is illustrated in Appendix~\ref{app:sed} (Fig.~\ref{fig:dsfr}), where a significative fraction of the galaxies with $\log(\sSFR_{tot}/\invYr)\le -10$  shows a much lower sSFR with the SED fitting method, leading to a lower $L_\txn{dust}$. 

In conclusion, the \NRK\ and SED fitting methods provide two independent and reliable estimates of $\Lir$ over a large range of redshift and galaxy physical parameters for the vast majority ($\sim 90$ \%) of the 24\,\micron\ star-forming sample. However, differences arise when considering the extreme sides of the population, such as highly star-forming (i.e., $\SFR_{tot}\ge 100 \,\MsunYr$) or quiescent galaxies [i.e., $\log(\sSFR_{tot}/\invYr)\le -10$], which represent $\la$ 1 \% and 7 \% of the whole sample, respectively.  
 While the \NRK\ method suffers from a bias at low sSFR, it is not clear the amplitude of this effect, since there is no robust indicators of SFR in the low activity regime. In the next Section, we obtain an independent estimate of this bias for the quiescent population by directly comparing the SFRs estimated with the \NRK\ and SED fitting methods. 
%%
%%%%%%%%%%%%%%%%%%%%%%%%%%%%%%%
\subsection{Comparison of SFR estimates with NRK and SED fitting techniques }
\label{sec:dep2}
 Once calibrated with a Far-IR sample, the \NRK\  technique can be applied to any sample of galaxies. In this Section we compare the SFRs derived with the SED fitting and \NRK\ methods for two samples of galaxies, the 24\,\micron\ and a K-selected (down to $K\le 23.5$) samples. We restrict the analysis to galaxies with  stellar masses $\txn{M}\ga 2\ 10^9 \Msun$ and  redshift  $0.05\le z\le 1.3$. The $\SFRnrk$ is estimated by means of Equations~\eqref{eq:sfr}--\eqref{eq:irx}. 
 As discussed above, we found that this technique is better suited for active star-forming galaxies than the less active population (low sSFR). As shown in Fig.~\ref{fig:ssfr}, we exploit the capability of the \NUVrK\  diagram to separate galaxies with different sSFRs and we define a new vector $\NRKv_{ssfr}$,  perpendicular to the starburst attenuation vector  ($\NRKv_{ssfr}=\cos(54^{\circ}) \times (NUV-r) -\sin(54^{\circ})\times (r-K)$).  In Fig.~\ref{fig:ssfr}, we show constant values of  $\NRK_{ssfr}$  (i.e., the norm of the $\NRKv_{ssfr}$ vector), corresponding to the range  $-2 \le NRK_{ssfr} \le 3$ as gray lines, which is a good proxy to follow the variation of the mean sSFR.  The limit adopted to define the passive population corresponds to $NRK_{ssfr}\ga1.9$.

\begin{figure}
	\centering
\includegraphics[width=\hsize]{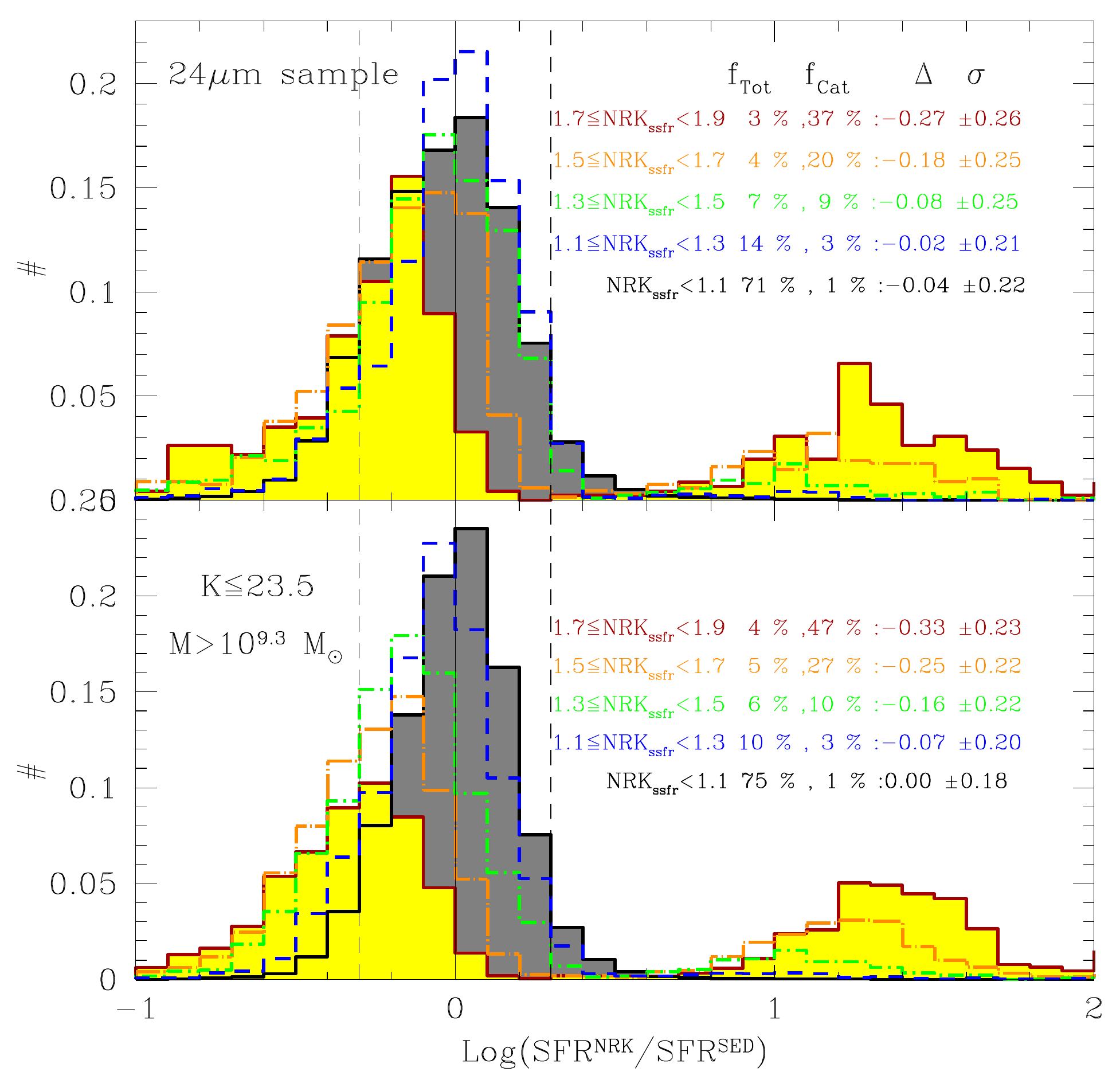}
\caption[]{ Distributions of the SFR ratios ($\SFRnrk/\SFRsed$) in different bins of  \NRK$_{ssfr}$. The distribution are renormalized to unity.  The top panel refers to the 24\,\micron\ sample while the bottom panel includes all galaxies with  $K\le 23.5$ and more massive than $\txn{M}\sim 2 \ 10^9 \,\Msun$. For each interval of $NRK_{ssfr}$, we report the fraction of objects ($f_{Tot}$) , catastrophic failures ($f_{Cat}$), the median ($\Delta$) and dispersion ($\sigma$).  }
\label{fig:dsfr_nrkp}
\end{figure}

%
%The differences  $(SFR_{NRK} - SFR_{SED})$ for the 24\,\micron\ (top panel) 
The ratios $(\SFRnrk /\SFRsed)$  for  the 24\,\micron\ (top panel)  
 and  the K (bottom panel) selected samples are shown in Fig.~\ref{fig:dsfr_nrkp}  for different intervals of  $\NRK_{ssfr}$. In each panel, we report for each  bin of $NRK_{ssfr}$  the percentage of objects in this bin,  the percentage of catastrophic objects (estimated as ABS$(\Delta)>3\sigma$), the median ($\Delta$) and scatter ($\sigma$).    
  For  $\sim 85$ \%  of the sample, with $NRK_{ssfr}\le 1.3$, the two SFRs  are in excellent agreement, with small biases, scatters and catastrophic fraction for both the 24\,\micron\ and K-selected samples.
For $1.3\le NRK_{ssfr}\le 1.5$ ($\sim6-7$ \% of  the two samples),  the fraction of catastrophic objects sharply increases to $\sim10$ \%, the dispersion also increases and the median starts to deviate from zero, with $\SFRsed$ being higher than $\SFRnrk$.  This trend becomes more severe for  $1.5\le NRK_{ssfr}\le 1.9$ (the remaining $\sim 7$ to 9 \% of the samples). The two SFRs disagree,  with a large fraction of catastrophic objects ($\ga 25$ \%), a significant non-zero median and large dispersion, with some slight differences between the two samples. This region with inconsistent SFR estimates is shown in Fig.~\ref{fig:ssfr} as a shaded area. 
 
This comparison shows that the \NRK\ method can be successfully applied to a sample of galaxies larger than the 24\,\micron\ sample, since it provides SFR estimates in good agreement with the SED fitting for the vast majority (i.e., $\sim 85-90$ \%) of the star-forming population.
 However, the methods diverge when including low-activity (low-sSFR) galaxies \citep[see also][]{Johnson2007, Treyer2007}.  To alleviate this limitation, we show that the population with discording results ($\le10$ \% of the whole sample) can be easily isolated in the \NUVrK\ diagram. 
%
%%%%%%%%%%%%%%%%%%%%%%%
%
%%%%%%%%%%%%%%%%%%%%%%%%%%%%%%%%%%%%%%%%%
%
\subsection{Application to the  SFR vs stellar Mass relationship }
\begin{figure}
	\centering
\includegraphics[width=\hsize]{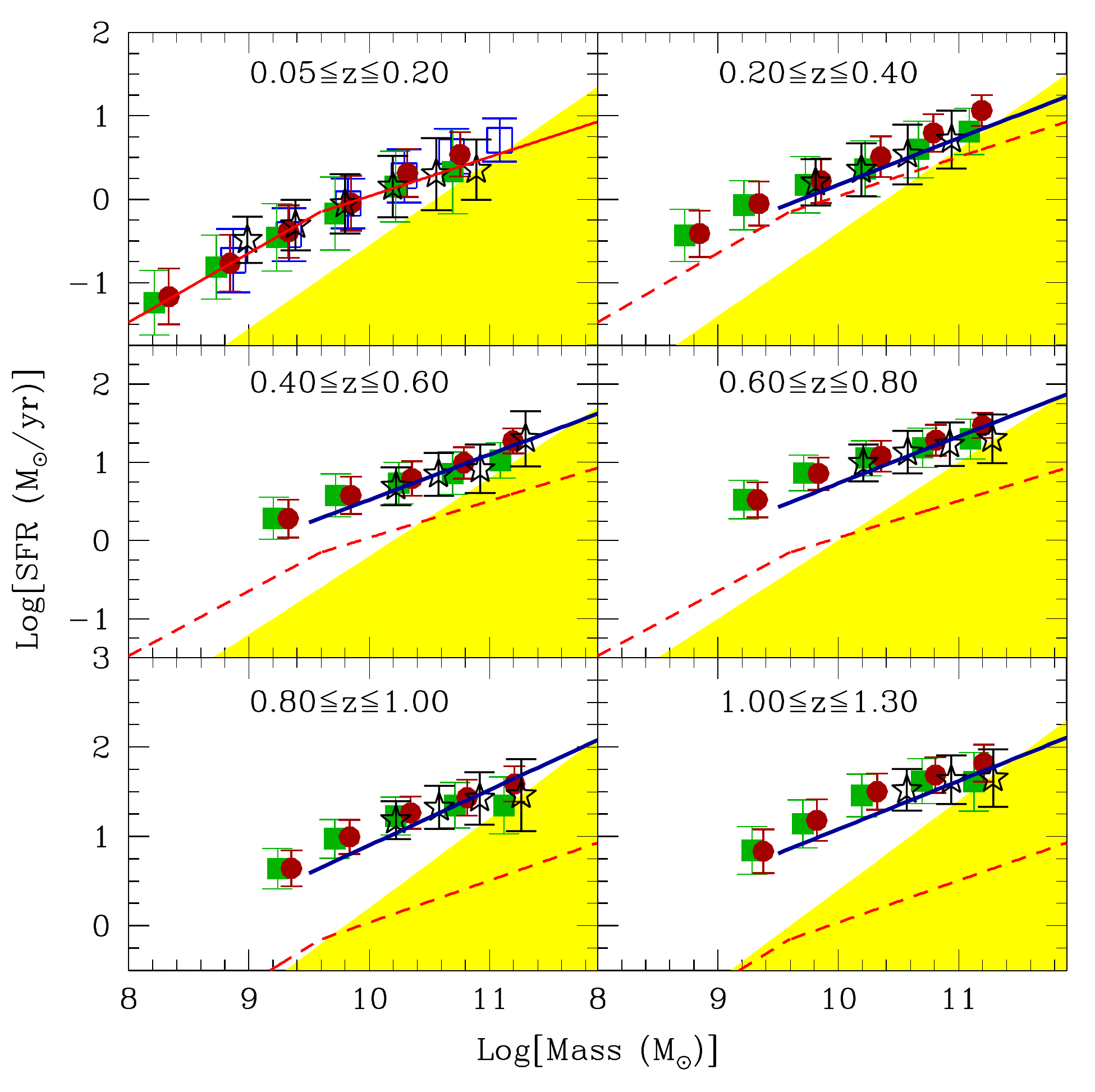} 
\caption[]{The SFR vs stellar Mass relation for star-forming galaxies in different redshift bins. The $V_{max}$ weighted means are shown for the 24\,\micron\ sample  (SFR based on Eq.~\ref{eq:sfr}, open black stars);  the COSMOS K-selected sample using the SFR derived from the \NRK\ approaches (original method: filled red circles and with sSFR-correction: filled green squares);  the local GALEX-SDSS-SWIRE sample (blue open squares). We compare with the local estimate from \citet{Salim2007}  (red lines) and the radio stacking analysis by \citet{Karim2011} (heavy dark-blue lines). The region where the \NRK\ method becomes less reliable is shown as light shaded area.  }
\label{fig:sfrm}
\end{figure}
 A correlation between SFR and stellar mass in star-forming galaxies is observed both at low \citep{Salim2007} and  high redshift \citep{Noeske2007a, Elbaz2007}. This relation is surprisingly tight, with an intrinsic scatter of $\sim 0.3$ dex.  We can therefore exploit the presence of this correlation and test the ability of the \NRK\ method to reproduce the slope and amplitude of this relation at different redshifts. 

In Fig.~\ref{fig:sfrm} we show the $V_{max}$ weighted mean of SFR in bins of stellar mass for different samples of star-forming galaxies: the COSMOS 24\,\micron\ sample, a flux limited K-selected sample ($K\le 23.5$) and, in the lowest redshift bin, the local GALEX-SWIRE sample.
Symbols indicate SFR estimated with different methods: $(L_{IR}+L_{UV})$ from Equation~\ref{eq:sfr} (black stars), \NRK\ method (red filled circles and open blue squares), \NRK\ method corrected for the sSFR bias (green-filled squares).

 At low-z, the different SFR estimates provide similar results and the SFR--\Mstar\ relation is in excellent agreement with the one measured by \citet{Salim2007} for the GALEX-SDSS sample and derived with a SED fitting method (solid red line).  At higher redshifts, we compare our finding with the results of \citet{Karim2011}, which are based on the 1.4 GHz radio continuum emission of stacked star-forming galaxies in different stellar mass and redshift bins (solid blue lines). We consider  their results from the two-parameter fits given in their Table 4. The SFR--\Mstar\ relation derived from individual galaxies with both the \NRK\ sSFR bias-uncorrected and corrected methods agrees well with the radio stacking from \citet{Karim2011}. 
  The major difference between the \NRK\  and the \NRK-sSFR corrected method is in the dispersions around the mean SFRs. The original \NRK\ method shows a small dispersion in the high mass end of the relation, due to an over-estimate of the SFR for massive, evolved galaxies, artificially moving them towards the SFR--\Mstar\ sequence. This effect varies with redshift, as described in Appendix~\ref{app:dep_nrk}, thus we indicate with the shaded area in Fig.~\ref{fig:sfrm} the regions in which the SFR is over-estimated by a factor greater than two. It can be seen that, at all redshifts, the bin corresponding to the most massive galaxies is affected by this bias, hence producing a smaller dispersion on the SFR--\Mstar\ relation with respect to that observed with the 24\,\micron\ and \NRK-sSFR corrected methods. We note that this effect becomes irrelevant at lower masses. 
 
 %
  %, this effect does not we do not expect this effect to play a significant role.  \\
  Overall, even if the stellar mass does not enter in the calibration of the \NRK\ method, the SFR estimated with this method for individual galaxies can reproduce the slope and normalisation of the SFR--\Mstar\ relation, along with its redshift evolution.   
 %
  
%
%%%%%%%%%%%%%%%%%%%%%%%%%%%%%%%%
\section{Modeling the \irx\  in the \NUVrK\ diagram}
\label{sec:discu}
%\subsection{Can models reproduce the observed trends?}
\label{sec:models}

In the previous Section~we have shown that galaxies with different ultraviolet-to-infrared luminosity ratios ($IRX$) are well separated in the \NUVrK\ diagram.
% This allows us to estimate the total IR luminosity from a combination of ultraviolet, optical and near-IR observations, which are easily accessible in a broad range of redshift.   
To confirm the validity of this approach, and to explain the physical origin of the observed trends, we appeal to a library of galaxy SEDs computed with the BC03 spectral evolution model. 
We follow the approach of \citet{Pacifici2012} and extract a set of star formation and chemical enrichment histories from the semi-analytic post-treatment of \citet{DeLucia2007} of the Millennium cosmological simulation \citep{Springel2005}. The star formation and chemical enrichment histories computed in this way reproduce the mean properties of nearby SDSS galaxies. Thus, they span only limited ranges in sSFR, around $10^{-10} \, \invYr$, and in the fraction of the current stellar mass formed in the last 2.5 Gyr at $z=0$. 
Following \citet{Pacifici2012}, to account for the broader range of spectral properties of the galaxies in our sample with respect to the SDSS, we re-draw the evolutionary stage at which a galaxy is looked at in the library of star formation and chemical enrichment histories (we do this uniformly in redshift between 0.2 and 1.5). We also resample the current (i.e., averaged over the last 30 Myr) SFR from a Gaussian distribution centered on $\log(\txn{sSFR})=-9.1$, with a dispersion of 0.6. This choice of parameters reproduces the observed global distribution of sSFR (i.e., summed over all redshift bins) and the distribution of galaxies in the \NUVrK\ diagram, after accounting for dust attenuation as described in the next Section.
We adopt the \citet{Chabrier2003} IMF. 

\subsection{Dust attenuation model}
To include the effect of dust attenuation, we adopt the dust prescription of \citet{Chevallard2013}. This extends the two-components, angle-average dust model of \citet{Charlot2000} to include the effect of galaxy inclination and different spatial distributions of dust and stars on the observed spectral energy distributions of galaxies. \citet{Chevallard2013} combine the radiative transfer model of \citet[][hereafter T04]{Tuffs2004} with the BC03 spectral evolution model. To accomplish this, they relate the different geometric components of the T04 model (a thick and thin stellar disks, and a bulge, attenuated by two dust disks) to stars in different age ranges. Here, for the sake of simplicity and to limit the number of adjustable parameters, we describe attenuation in the diffuse ISM using only the thin stellar disk model of \citet{Tuffs2004}. This is supported by the finding by \citet{Chevallard2013} that, in a large sample of nearby star-forming galaxies, the thin stellar disk component of the T04 model accounts for $\approx 80$ percent of the attenuation in the diffuse ISM. Also, we note that adding the T04 thick disk component has a weak effect on the results. The dust content of the diffuse ISM in the T04 model is parametrized by means of the $B$-band central face-on optical depth of the dust disks \taubP . This determines, at fixed geometry, the attenuation of starlight by dust at any galaxy inclination $\theta$, which measures the angle between the observer line-of-sight and the normal to the equatorial plane of a galaxy. At fixed \taubP, the integration over the solid angle of the attenuation curve \tauLismTh \ in the T04 model yields the angle-average attenuation curve \meantaulISM\ \citep[see Section~2 of][]{Chevallard2013}. 
As in \citet{Charlot2000}, we couple the attenuation in the diffuse ISM described by the T04 model with a component describing the enhanced attenuation of newly born stars ($t<10$ Myr) in their parent molecular clouds. Following \citet{Charlot2000}, we parametrize this enhanced attenuation by means of the fraction $1-\mu$ of the total attenuation that arises from dust in stellar birth clouds, in the angle-average case. 

The attenuation of the radiation emitted by a stellar generation of age $t$ at inclination $\theta$ can therefore be written as
\begin{equation}
\hat{\tau}^\txn{tot}_\lambda(\theta,t)=\left\{ \begin{array}{l l}
\tauLbc + \tauLismTh & \hspace{3mm} \txn{for} \hspace{3mm} t \leqslant 10\,\txn{Myr}\,,\\
\tauLismTh & \hspace{3mm} \txn{for} \hspace{3mm}  t>10\,\txn{Myr} \, , \end{array}\right.
\label{eq:tau_cf00}
\end{equation}
where the superscripts `BC' and `ISM' refer to attenuation in the birth clouds (assumed isotropic) and the diffuse ISM, respectively. In this expression the attenuation curve for the diffuse ISM \tauLismTh\ is taken from the T04 thin stellar disk model, and following \citet[see also \citealt{daCunha2008}]{Wild2007}, we compute the attenuation curve in the birth clouds as
\begin{equation}
\tauLbc = \tauVbc \, \left (\lambda / 0.55\,\micron \right ) ^ {-1.3}\, ,
\end{equation}
where the $V$-band optical depth of the birth clouds \tauVbc\ is related to the angle-average optical depth of the diffuse ISM \meantauVISM\ as
\begin{equation}
\tauVbc = (1-\mu) / \mu \, \meantauVISM \, .
\end{equation}

To compute the ratio of the infrared-to-ultraviolet luminosities $IRX$ in this model, we take the IR luminosity to be equal to the fraction of all photons emitted in the range $912\,\txn{\AA}\le~\lambda~\le~3\,\micron$ in any direction that are absorbed by dust (dust is almost transparent at $\lambda > 3$ \micron). Assuming that photons at IR wavelengths emerge isotropically from a galaxy, we write the IR luminosity $\Lir$ as 
\begin{equation}
\Lir = \frac{1}{4\pi} \int_\Omega d\Omega \int_{0.0912}^3 d\lambda [1-\exp({-\tauLTh})] L^0_{\lambda} \,
\end{equation}
where $L^0_{\lambda}$ is the unattenuated luminosity emitted by stars (assumed isotropic), $\Omega$ is the solid angle, and \tauLTh\ is the integral of equation~\ref{eq:tau_cf00} over the star formation history of the galaxy.
We compute the monochromatic ultraviolet luminosity at the frequency $\nu$ corresponding to $\lambda = 2300$\,\AA\ as ${\cal L}_{NUV}(\theta)=\nu L_{\nu}(\theta)$, where $L_{\nu}(\theta)$ is given by
\begin{equation}
L_\nu(\theta) = \left [ \exp{(-\tauLTh)} \right ] L^0_{\lambda} \, \frac{\lambda^2}{c} \, ,
\end{equation}
and $L^0_{\lambda}$ if the luminosity emitted by all stars at $\lambda = 2300$ \AA\ in the direction $\theta$, and $c$ is the speed of light.
\begin{figure}
\centering
\includegraphics[width=\hsize]{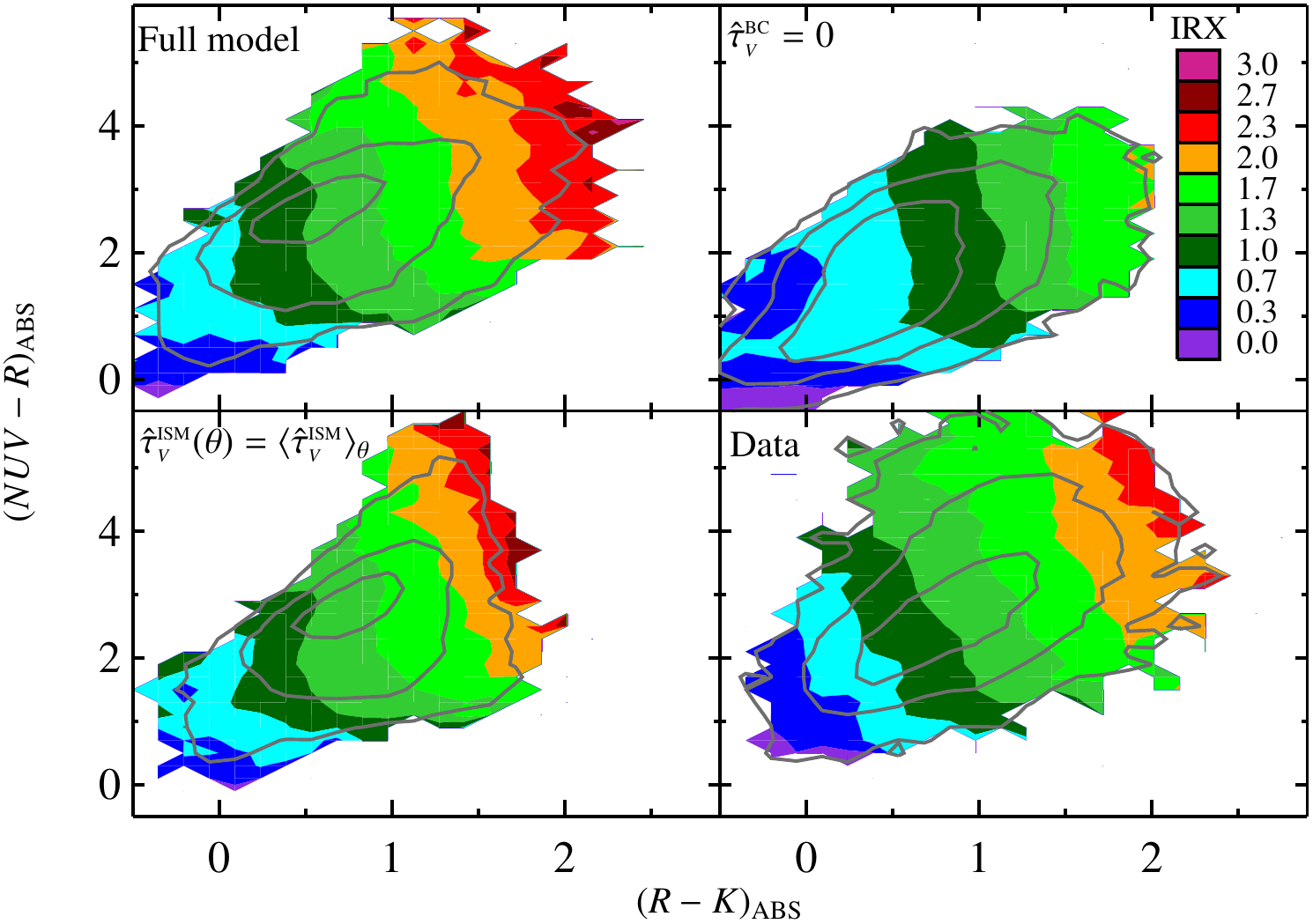}  \\
\caption[]{ 
Values of \irx\ (color coded on a logarithmic scale) in the \NUVrK\ diagram. The solid gray lines in each panel indicate the number density contour of the galaxies corresponding to 0.01, 0.1 and 0.5 the maximum density. Top-left panel, 20\,000 model SEDs computed with the `full model' (see Section~\ref{sec:models}). This includes the dust prescription of \citet{Chevallard2013}, which accounts for the effect on dust attenuation of galaxy geometry, inclination and enhanced attenuation of young stars by their birth clouds. Top-right panel, same as top-left panel, but neglecting the enhanced attenuation of young stars, i.e., fixing the birth clouds optical depth $\tauVbc = 0$. Bottom-left panel, same as top-left panel, but neglecting the effect of galaxy inclination, i.e., adopting the angle-averaged attenuation curves \meantaulISM . Bottom-right panel, data (see Section~\ref{sec:infrared_exc} and Fig.~\ref{fig:irx}).
}
\label{fig:mod}
\end{figure}

\subsection{Model library}
\begin{figure}
\centering
\includegraphics[width=\hsize]{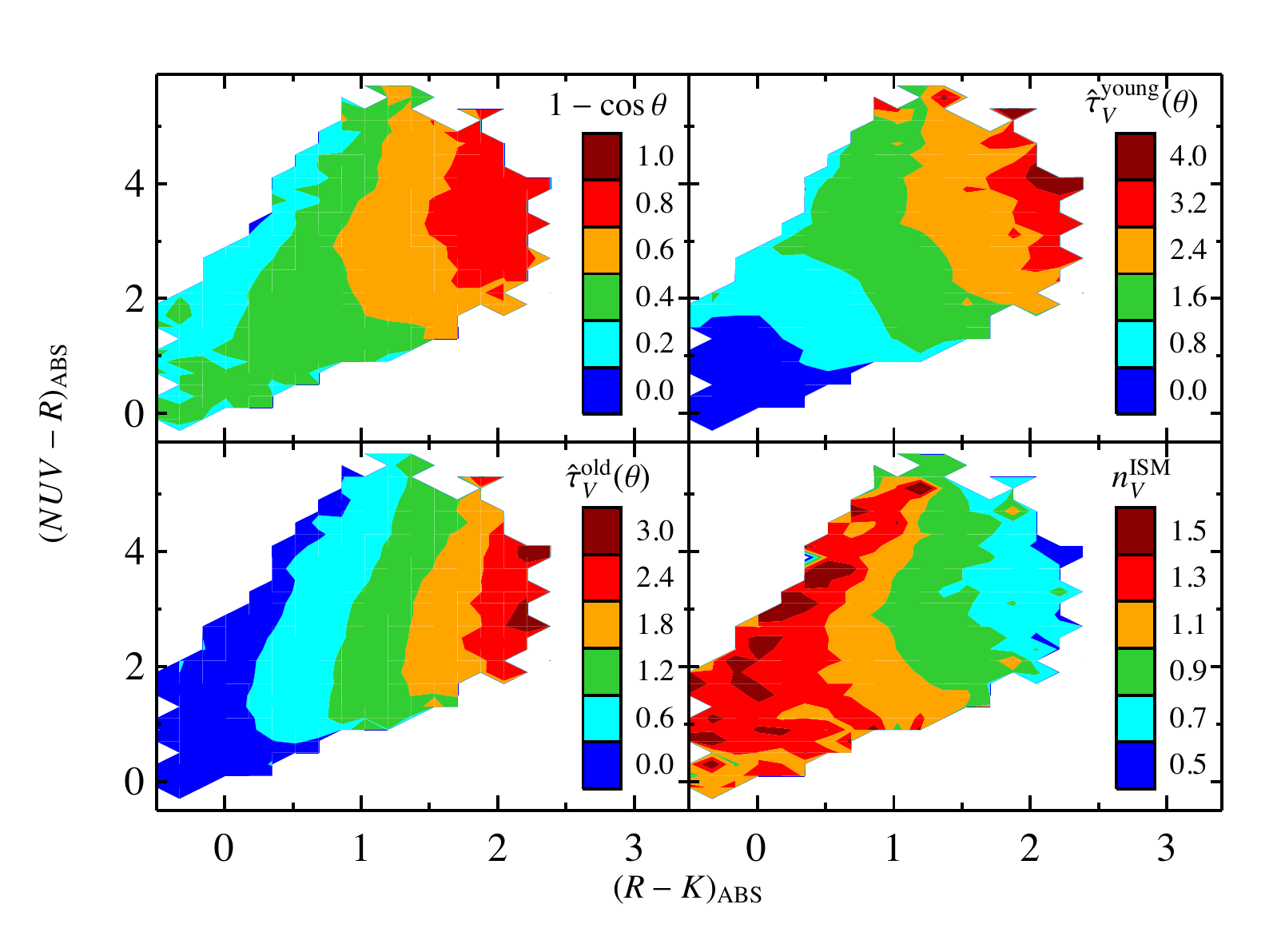}  \\
\caption[]{Mean value in bins of constant $(NUV-r)$ and $(r-K)$ of different parameters describing attenuation of starlight from dust for the model SEDs described in Section~\ref{sec:models}. Top-left panel: galaxy inclination $1-\cos\theta$. Top-right panel, $V$-band attenuation optical depth suffered by stars younger than $10^7$ yr \tauVyoung. Bottom-left panel, $V$-band attenuation optical depth suffered by stars older than $10^7$ yr \tauVold. Bottom-right panel, slope of the optical attenuation curve in the diffuse ISM, \nVism, measured from a power law fit to the model attenuation curves in the range $0.4 \leq \lambda \leq 0.7$ \micron.  
}
\label{fig:mod2}
\end{figure}

We use this model to compute a library of 20\,000 SEDs of dusty star-forming galaxies, which we divide in bins of constant $(NUV-r)$ and $(r-K)$. We compute the mean sSFR and \irx\  in each bin and explore their distribution in the \NUVrK\ diagram. After some experimentation, we find that a Gaussian distribution of \taubP\ centered at 7, with a dispersion of 3, truncated at the maximum value of available models $\taubP = 8$, and a Gaussian distribution of \,$\mu$ centered at 0.3, with a dispersion of 0.2, and truncated at \,$\mu = 0$ and \,$\mu=1$, allow us to well reproduce the data, as shown in the top-left panel of  Fig.~\ref{fig:mod}. We note that a uniform distribution of galaxy inclinations would produce a large tail of highly attenuated galaxies, which is not observed in the data (at $(r-K)> 2.5$). Hence, in Fig.~\ref{fig:mod} we have adopted the observed distribution of axis ratios, converted to inclinations using the standard formula for an oblate spheroid \citep[e.g.,][]{Guthrie1992}
\begin{equation}
\cos{\theta} = \sqrt{\frac{(b/a)^2 - q_0^2}{1-q_0^2}} \, ,
\end{equation}
where $q_0$ is the intrinsic axis ratio of the ellipsoid representing the galaxy, which we fix to $q_0=0.15$.

A comparison between top-left and bottom-right panels of Fig.~\ref{fig:mod} shows that the models reproduce, at least qualitatively, the distribution of \irx\ in the color-color plane.  
  The reddest near-IR colors, $(r-K)> 1.4$, correspond to galaxies seen at large inclination. This is consistent with figure \ref{fig:ell}, which shows that the galaxies with reddest $(r-K)$ colors have the largest measured ellipticities, i.e., they are more inclined. A large inclination makes the disk appear more opaque, since photons have to cross a larger section of the dust disk before they escape toward the observer.

The location and shape of the $IRX$ stripes in the theoretical \NUVrK\ diagram depend on several galaxy physical parameters, namely evolutionary stage, current SFR, dust content and distribution. The evolutionary stage and current SFR determine the relative amount of young and old stars in the galaxy, which controls the ratio of unattenuated ultraviolet to optical and near-IR luminosity of the galaxy. The global dust content, \tauLbc + \tauLismTh, and the distribution of dust between ambient ISM and birth clouds affect the $(NUV-r)$ and the $(r-K)$ colors in different ways. For galaxies with a non-negligible fraction of young stars (i.e., $\log(\txn{sSFR})\ga -9$), the $(NUV-r)$ color is mainly driven by the  stellar birth clouds optical depth \tauLbc, and the $(r-K)$ color by the optical depth of the diffuse ISM \tauLismTh.
  
We test the effect of varying the optical depth of stellar birth clouds by computing the same library of 20\,000 galaxy SEDs as described above, but fixing $\tauVbc=0$. Top-right panel of Fig.~\ref{fig:mod} shows that neglecting birth clouds attenuation prevents us from reproducing the reddest $(NUV-r)$ color observed in the data. The stripes appear almost perpendicular to the $(r-K)$ color driven primarily by the diffuse ISM.  Also, this model predicts smaller values of \irx\ at a fixed position in the $(NUV-r)$ vs $(r-K)$ diagram, since the UV photons emitted by young stars do not suffer enhanced attenuation by the dusty birth cloud environment, which would be re-emitted at IR wavelengths increasing the overall IR luminosity. 

We also study the effect of neglecting the dependence of dust attenuation on galaxy inclination. To achieve this, we compute the same library of 20\,000 SEDs as above, but we fix the attenuation curve to the angle-averaged curve \meantaulISM . Bottom-left panel of Fig.~\ref{fig:mod} shows that this prevents us from reproducing the reddest $(r-K)$ colors of the observed galaxies, which correspond to highly inclined objects (see top-left panel of Fig.~\ref{fig:mod2} and Fig.~\ref{fig:ell}).

We have shown in Fig.~\ref{fig:mod} that to reproduce qualitatively the observed distribution of galaxies and the value and orientation of the \irx\ stripes in the $(NUV-r)$ vs $(r-K)$ plane we need a prescription for dust attenuation which includes both a two-component medium (i.e., ISM + birth clouds) and the effect of galaxy inclination. We can now consider the `Full model' shown in the top-left panel of Fig.~\ref{fig:mod} and study how different dust properties vary in the $(NUV-r)$ vs $(r-K)$ plane. Fig.~\ref{fig:mod2} shows the same library of SEDs as in the top-left panel of Fig.~\ref{fig:mod}. As for Fig.~\ref{fig:mod}, we divide the galaxy SEDs in bins of constant $(NUV-r)$ and $(r-K)$, and compute in each bin the mean value of galaxy inclination $1-\cos\theta$, $V$-band attenuation optical depth seen by stars younger [older] than $10^7$ yr \tauVyoung\ [\tauVold], and slope of the optical attenuation curve in the diffuse ISM \nVism.

The top-left panel of Fig.~\ref{fig:mod2} shows that the galaxy inclination systematically increases as $(r-K)$ increases, with a weaker dependence on the $(NUV-r)$ color. This can be understood in terms of the attenuation optical depth in the diffuse ISM, which increases with increasing galaxy inclination, hence making $(r-K)$ redder. The $(NUV-r)$ color is also influenced by the variation of galaxy inclination, but to a much less extent since it also depends on the birth clouds attenuation optical depth [see Eq.~\eqref{eq:tau_cf00}].
The variation of the mean galaxy inclination shown in the top-left panel of Fig.~\ref{fig:mod2} is also in qualitative agreement with Fig.~\ref{fig:ell}, which shows that the mean observed ellipticity of the galaxies in our sample increases from the bottom-left to the top-right side of the $(NUV-r)$ vs $(r-K)$ plane.

The top-right panel of Fig.~\ref{fig:mod2} shows the variation of the $V$-band attenuation optical depth suffered by stars younger than $10^7$ yr [i.e., \tauVyoung, see Eq.~\eqref{eq:tau_cf00}]. At small $(r-K)$ the stripes of constant \tauVyoung\ are more parallel to the $(NUV-r)$ color, while they become more and more inclined as $(NUV-r)$ and $(r-K)$ increase. This behavior can be understood in terms of the different fraction of light emitted by young stars attenuated by dust in the birth clouds and in the diffuse ISM. At small (i.e., blue) $(r-K)$, young stars are mostly attenuated by the birth clouds component, which makes the $(NUV-r)$ larger (i.e., redder, by decreasing $NUV$, at fixed $r$) without affecting $(r-K)$. As $(r-K)$ increases, the attenuation in the diffuse ISM increases, because galaxies are more inclined or have a larger dust content, and so the fraction of light emitted by young stars attenuated by this component raises too. As a consequence, the $(NUV-r)$ color is determined by attenuation in both components, and the stripes of constant \tauVyoung\ change orientation.

The variation of the $V$-band attenuation optical depth suffered by stars older than $10^7$ yr (i.e., \tauVold) shown in the bottom-left panel of Fig.~\ref{fig:mod2} follow that of the $(r-K)$ color, as indicated by the orientation of the stripes of constant \tauVold\ almost perpendicular to $(r-K)$. This is not surprising, since $(r-K)$ traces stars older than $10^7$ yr, which are attenuated by dust in diffuse ISM. When moving from left to right on the $(r-K)$ axis, \tauVold\ increases from ~0 to ~2.5, indicating that the amount of attenuation suffered by stars in galaxies with very large (i.e., red) $(r-K)$ is substantial.

The bottom-right panel of Fig.~\ref{fig:mod2} shows the slope of the optical attenuation curve in the diffuse ISM \nVism, obtained by fitting a power law to the model attenuation curves in the range $0.4 \leq \lambda \leq 0.7$ \micron. The slope \nVism\ becomes smaller (i.e., the attenuation curve becomes flatter) when moving from the bottom-left to the top-right side of the diagram. This effect, as described in \citet{Chevallard2013} (see their fig~4), is a general prediction of radiative transfer models which consider disk galaxies with a mixed distribution of dust and stars. The variation of the slope of the attenuation curve in the $(NUV-r)$ vs $(r-K)$ plane can account for the fact that the observed stripes of constant \irx\ are not perpendicular to the SMC and Calzetti attenuation vectors (see Fig.~\ref{fig:irx}) and that the SED fitting predicts a systematic variation of the slope of the attenuation curve as a function of sSFR (see Fig.~\ref{fig:ext_law}).

With this analysis we have shown that we must account for the effect of geometry (i.e., of the spatial distribution of dust and stars) and galaxy inclination to reproduce the attenuation of starlight by the diffuse ISM, which mainly affects the $(r-K)$ color. We have also shown the importance of accounting for the enhanced attenuation of newly born stars by their birth clouds to reproduce the reddest $(NUV-r)$ colors and to match the observed values of \irx . In the end, the good qualitative agreement between the "fully model" galaxies and the data (i.e., top-left and bottom-right panels of Fig.~\ref{fig:mod}) confirms that the \NUVrK\ diagram encodes valuable information about the global energy transfer between starlight and dust and galaxy inclination. We defer to a future work a more detailed and quantitative analysis of the data presented here, which would help us to better constrain the global amount, distribution and redshift evolution of the dust in star-forming galaxies.

%
%
%%%%%%%%%%%%%%%%%%%%%%%%%%%%%%%%%%%%%%%%
%
%
\section{Conclusion}

 We  present a new method  to compute the SFR of individual star-forming galaxies based on their location in the $(NUV-r)$ versus $(r-K)$ color-color diagram.  
 We show that  the  \NUVrK\ diagram  provides an efficient way to separate quiescent and star-forming galaxies, an alternative to the UVJ diagram proposed by \citet{Williams2009}. For the star-forming galaxies, the UV/optical luminosities in this diagram are highly sensitive to the shape of  the dust attenuation laws.  On the other hand, the infrared excess, $IRX=\Lir/\Luv$, as the net budget of the absorbed versus unabsorbed UV light, is weakly dependent on these effects.  
  We combine the two dust diagnostics by  analyzing the distribution of the mean infrared excess (i.e., \irx)  in the  \NUVrK\  diagram for a large sample of star-forming galaxies at redshift $0 \le z \le 1.3$ selected from the COSMOS 24\,\micron\ and low-z GALEX-SDSS-SWIRE \citep{Johnson2007} samples. 
  We observe the presence of stripes with constant \irx , associated with a small dispersion around the mean,  which allows us to describe \irx\ with a unique vector [i.e., \NRKv, a combination of $(NUV-r)$ and $(r-K)$ colors].  We derive a simple relation between \irx\ and \NRK, the norm of the vector \NRKv, and redshift, valid for star-forming galaxies with $\txn{M}\ge 2\ 10^{9}\, \Msun$ and $0.05-0.1\le z\le 1.3$. This relation allows us to predict the IR luminosity of individual galaxies with an accuracy of $\sim0.2$ dex (up to 0.27 dex for the local sample), which is better than the accuracy obtained with the SED fitting method based on 31 COSMOS pass-bands for the same galaxies.

We perform extensive comparisons of the $\Lir$ and SFRs derived with the 24\,\micron, \NRK\ and SED fitting methods. We find that the three methods provide consistent results for the vast majority of star-forming galaxies ($\sim 85-90$ \%). The methods diverge for highly star-forming galaxies ($\SFR\ge 100\,\MsunYr$), which remain a negligible population at $z\le 1.3$, and for more evolved galaxies ($\sSFR\la 10^{-10} \,\invYr$). For the latter, we  describe a sSFR-, redshift-dependent limit below which the \NRK\ method becomes unreliable and we also show that this population with inconsistent SFR estimates can be easily isolated and discarded in the \NUVrK\ diagram. 

 By using the \NRK\ method, we reconstruct  the relationship between SFR and stellar mass for a $K$-selected sample of star-forming galaxies and find an excellent agreement with previous results over the entire redshift range.

Finally, we investigate  the physical origin of the \irx\ stripes in the \NUVrK\ diagram by appealing to a library of model SEDs based on the population synthesis code of \citet{BC03}.
 We find that this library of models is able to qualitatively reproduce the location and shape of \irx\ stripes in the \NUVrK\ diagram if we adopt a realistic prescription for dust attenuation \citep{Chevallard2013}. We show that to reproduce the observed stripes of \irx\ we must appeal to a two-component (i.e., birth clouds + diffuse ISM) dust model, which must account also for the effect on dust attenuation of galaxy inclination and geometry (i.e., the spatial distribution of dust and stars).\\
 The  method discussed in this work offers a simple alternative to assess the total SFR of star-forming galaxies in the absence of Far-IR observations or  spectroscopic diagnostics. Because it directly predicts the infrared excess, no assumption on the dust attenuation curves  is required to derive the SFR, in contrast to other methods such as the $\beta$-slope or SED fitting. \\
 In a companion paper \citet{Lefloch2013}, we extend our analysis toward lower stellar mass and higher redshifts based on the stacking technique of the Far-IR emission using the complete dataset available from Spitzer/MIPS at 24\,\micron\ to Herschel/SPIRE at 250, 350 and 500 \micron.   
\acknowledgement{We thank the anonymous referee for his useful comments. This work has been partially supported by the CNRS-INSU and the Programme National de Cosmologie et Galaxies. JC acknowledges the support of the European Commission through the Marie Curie Initial Training Network ELIXIR under contract PITN-GA-2008-214227. }
%%%%%%%%%%%%%%%%%%%%%%%%%%%%%%%
%
\bibliography{bibtex_arnouts} 
%
%%%%%%%%%%%%%%%%%%%%%%%%%%%%%%%
\appendix
\section{The SED fitting technique}
\label{app:sed}
 The broad band  SED fitting technique is a simple approach to infer the
 physical properties of a galaxy, such as stellar mass, SFR,  amount of dust , age of stellar populations,  by statistically comparing model and observed SED.
 The constraints on the physical parameters depend on the wavelength range spanned by the data and their quality. The COSMOS field, for which a wealth of multi-wavelength, high S/N observations exist, is thus well suited for such modeling.
 
 To derive the  physical parameters, we adopt a library of spectral energy distributions based on the synthetic stellar population code from \citet[ hererafter BC03]{BC03}.
  We describe the star formation history either with an exponentially declining function, with e-folding time $0.01\le \tau \le15$ Gyr, or with a constant. We adopt  two metallicities, subsolar and solar (i.e., $Z_{\odot}$, $0.2\,Z_{\odot}$), and  the IMF of \citet{Chabrier2003}, truncated at 0.1 and 100 \Msun. Since the maximum redshift of the galaxies in our sample is $z ~ 1.3$, we force the age of galaxies to be larger than 100\,Myr, computed from the onset of SF which, in our case, corresponds to the initial burst.
  We also constrain the age of the galaxies not to exceed  the age of the universe at any redshift. 
  We do not adopt rising star formation histories, since these were developed to improve the SED fitting at high redshift \citep[$z\ge2$, see][]{Maraston2010}. 
  The prescriptions for TP-AGB stars adopted in BC03 produces a lower near-IR luminosity for intermediate age stellar populations with respect to the prescriptions of \citet{Maraston2005}. This affects the galaxy mass-to-light ratio, and produces a difference in the stellar mass estimated with \citet{Maraston2010} of $\sim$-0.15 dex. 

 The dust attenuation curve encodes informations about the nature of the dust grains (sizes, chemical compositions) and the spatial distributions  of dust and stars.
  \citet{Boquien2009} have shown the necessity of adopting a range of attenuation laws to reproduce   the observed scatter in the IRX vs $\beta$ relation.
 In particular, they show that a gray (i.e., shallow) attenuation curve, such as the starburst curve of \citet{Calzetti2000}, and a steeper curve, such the SMC extinction curve of \citet{Prevot1984}, are required to span the observed distribution in the IRX versus $\beta$ diagram of the starbursting and normal star-forming galaxies. Similar conclusions are reached by
 \citet{Ilbert2009}, who find that a range of attenuation curves is required to improve the photometric redshift accuracy via the SED fitting method. For these reasons, we adopt three different attenuation curves:  a starburst, SMC-like laws and a curve with a slope in between them ($\lambda^{-0.9}$). We then consider reddening excess  in the range $0\le E(B-V)\le 0.6$, which allows us to explore the observed   color distribution of our sample (see Fig.~\ref{fig:nuvrk}).  
 
  We use Le Phare code \citep{Arnouts1999, Ilbert2006}  to compute the $\chi^2$  for each observed galaxy and the entire model library,  with all the photometric  passbands from 0.15\,\micron\ to 4.5\,\micron\ . The physical parameters are derived by computing the median of the marginalized likelihood for each parameter and the errors corresponding to the 68 \% credible region.
%          
%%%%%%%%%%%%%%%%%%%%%%%%%%%%%%%%%
\begin{figure}
%\begin{tabular}{c}
	\centering
\subfigure{ \includegraphics[width=\hsize]{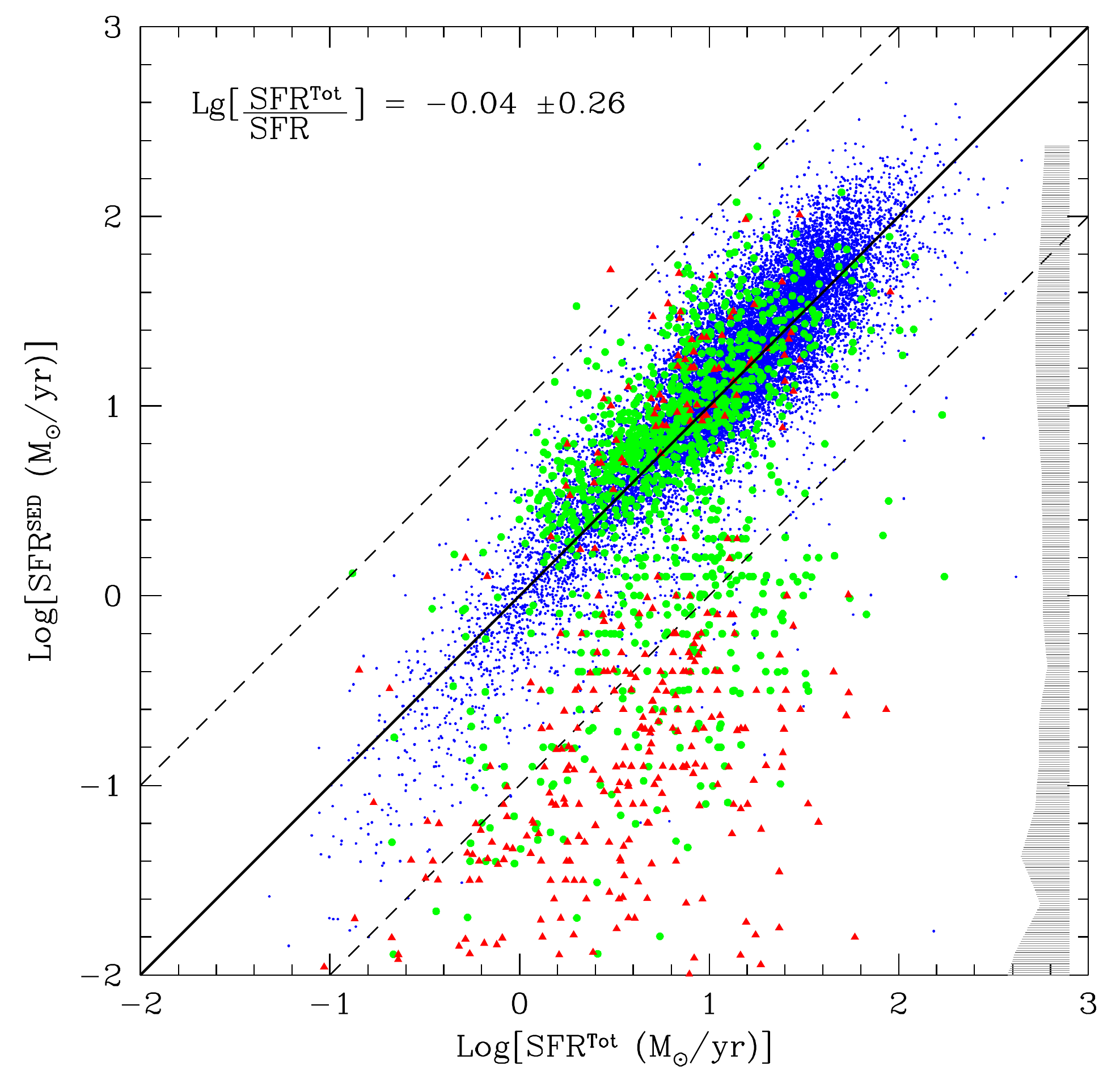} } 
\subfigure{ \includegraphics[width=\hsize]{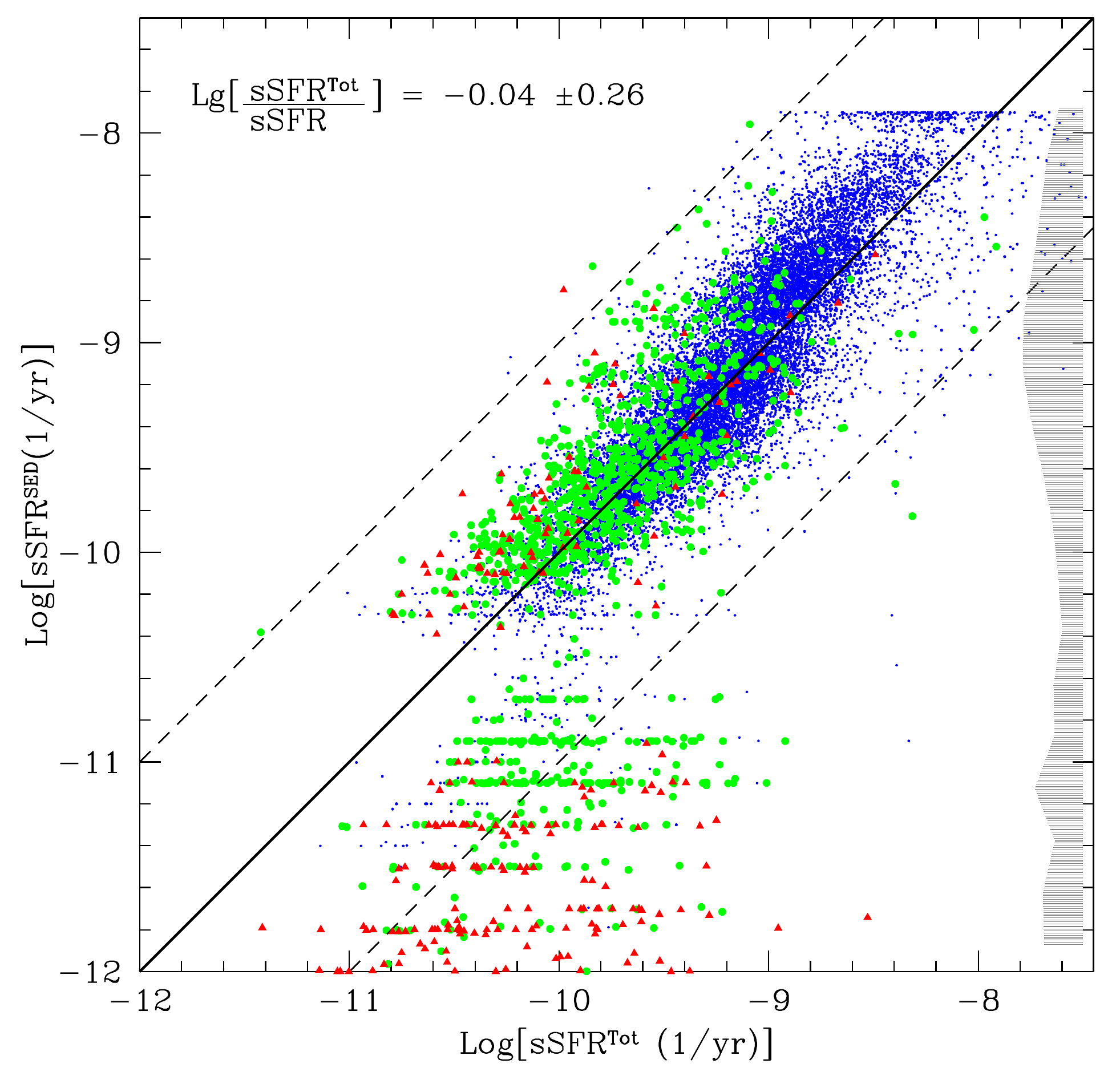} }
%\end{tabular}
\caption[]{ Comparison between the  SFR (top panel) and sSFR (bottom panel)  derived from the  SED fitting and            the IR+UV method  (i.e., Equation~\ref{eq:sfr}).  The mean errors on the SFR and sSFR estimated  are shown as shaded region in the right side of the plots.  The  passive galaxies and the ones in the "intermediate" region of \NUVrK\  diagram are shown as red triangles and green circles respectively. The  mean and dispersion of the relation reported in each figure do not include the passive galaxies. }
\label{fig:dsfr}
\end{figure}
%%%%%%%%%%%%%%%%%%%%%%%%%%%%%%%%
%

 In Fig.~\ref{fig:dsfr}, we compare the instantaneous SFR and sSFR derived with the SED fitting with the "total" SFR and sSFR derived from the observed IR and UV luminosities (i.e., Equation~\ref{eq:sfr}). The mean errorbars (based on 68 \% errors) for the SED parameters vary between 0.2 to 0.4,  as shown by the gray region on the right side of the figures.   

 For the bulk of the 24\,\micron\ population the SFRs are in good agreement over $\sim$3 order of magnitude, with a dispersion lower then a factor of two. The  vast majority of sources with large discrepancies is located in the region occupied by passive galaxies (red triangles) or next to it,
  in the "intermediate" zone (green dots) as defined in Fig.~\ref{fig:ssfr}. For those galaxies, the SED fitting predicts a low specific SFR (i.e., $\sSFRsed\le 10^{-10.5} \, \invYr$). The origin of this difference may have multiple causes: an inadequate description of dust attenuation may cause the SED fitting to reject highly attenuated models and prefers models with low or no dust content and a low specific SFR.  Alternatively the adopted definition of the total SFR in Eq~\ref{eq:sfr}  may over-estimate the SFR, since we neglect the contribution of old stars to the dust heating  ($(1-\eta) \Lir$). This contribution is often  considered as 30 \% ($\eta=0.3$) for star-forming galaxies \citep{Inoue2005} but it can be higher for the most evolved galaxies  \citep[e.g.,][]{Cortese2009}.  However,  even an extreme value of $\eta\sim 0.9$  will not  reconcile the  estimated SFRs with the two methods. Another possibility is that the extrapolation of the 24\,\micron\ flux into the total IR luminosity could fail for galaxies with  low specific SFR, if for example, a warmer dust temperature is associated to the same 24\,\micron\ flux with respect to galaxies in the star-forming main sequence, as recently reported by \citet{Skibba2011, Smith2012}. It is beyond the scoop of this paper to address this issue, since we focus on the star-forming galaxies, but we find that $\sim 7$ \% of the entire 24\,\micron\ sample is affected by this mismatch in the SFR estimates.
    
\begin{figure}
	\centering
\includegraphics[width=\hsize]{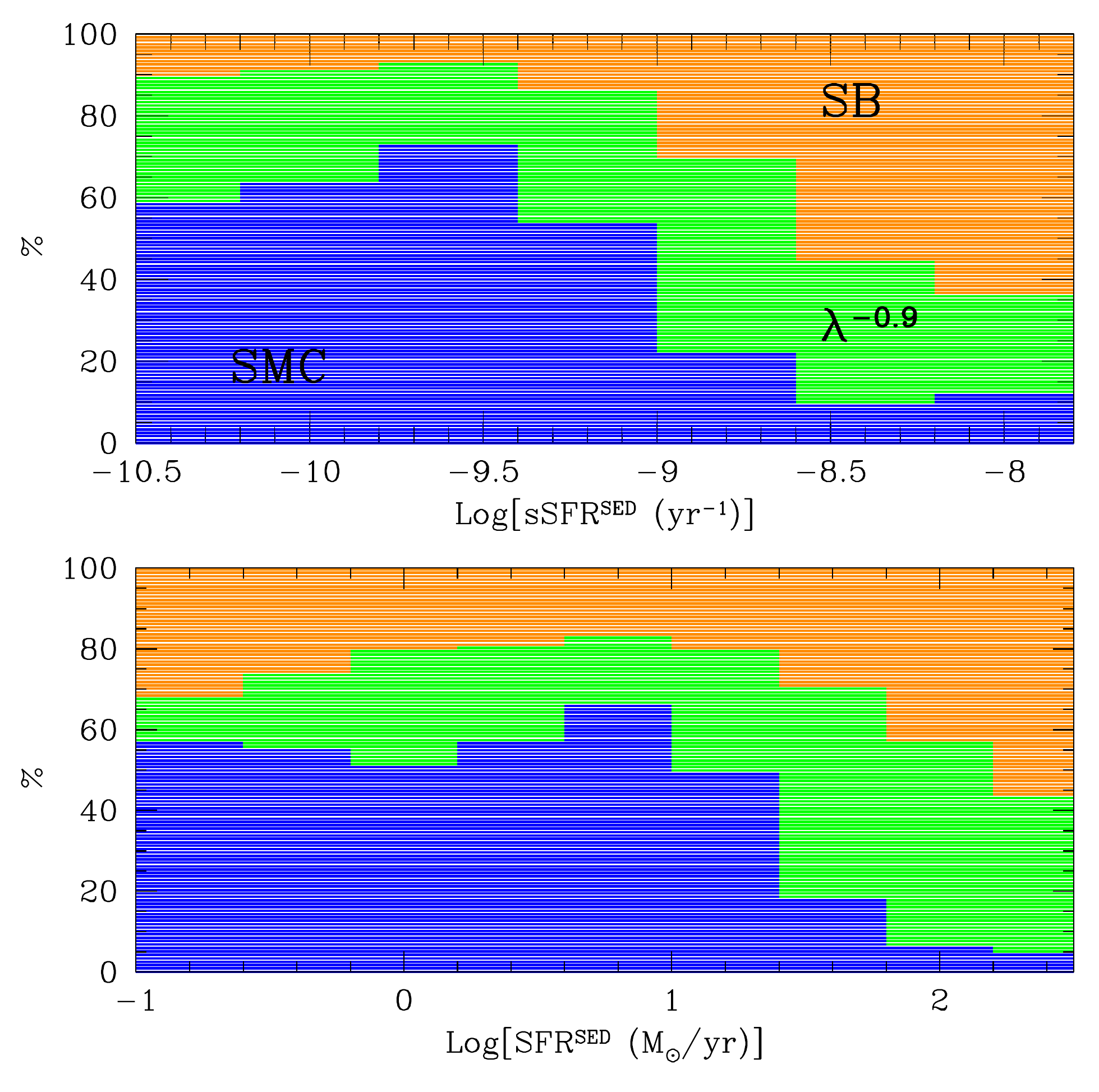} 
\caption[]{Relative contribution of the three attenuation laws (starburst [orange]; SMC [blue] and intermediate, $\propto\lambda^{-0.9}$ [green]) as a function of the specific SFR (top figure) and SFR (bottom figure) and derived from the SED modeling. 
    }
\label{fig:ext_law}
\end{figure}

  In Fig.~\ref{fig:ext_law}, we show the relative contribution of the different attenuation laws, corresponding to the best-fit templates, as a function of the SFR and specific SFR.  As mentioned above,  the SMC-like extinction law is favored for galaxies with low SFR and/or low sSFR while the starburst law better fits the active/starbursting galaxies with high SFR and sSFR ($\sSFR \ge -9$).  Our results agree with  \citet{Wuyts2011} with a  transition for a steeper attenuation law  at  $\SFR\le 20 \,\MsunYr$.   The most active galaxies are  consistent with a  mixed distribution of the dust and star resulting in the gray attenuation law \citep{Calzetti2000},  while "normal" star-forming galaxies  prefer  the SMC-like attenuation consistent with a simple dust screen model.   See section~\ref{sec:discu} and  \citet{Chevallard2013} for a purely geometric origin of different attenuation laws.
 
In conclusion, our analysis shows that SFRs estimated via the SED fitting technique are in good agreement with those derived from the UV+IR contribution. This validates our method to derive the $\Lir$ from the 24\,\micron\ flux and the use of Eq~\ref{eq:sfr} as a good measure of the SFR. We have noted that Equation~\eqref{eq:sfr} may become inadequate to describe the SFR in more evolved galaxies, possibly because of the presence of a larger population of old stars or the inadequate conversion of $L_{24\mu m}$ to $\Lir$.
 While this problem can not be easily solved, in this paper we show that we can isolate in the \NUVrK\ diagram the region occupied by galaxies  for which we obtain inconsistent SFR estimates with the different methods.
%  
%%%%%%%%%%%%%%%%%%%%%%%%%%%%%%%%
%%%%%%%%%%%%%%%%%%%%%%%%%%%%%%%%
%  
\section{Separation between Passive and active}
\label{app:passive}
%%  
%%%%%%%%%%%%%%%%%%%%%%%%%%%%%%%
%%  
 As the UVJ diagram proposed by \citet{Williams2009}, in Section~\ref{sec:nuvrk} and in Fig.~\ref{fig:ssfr}, we define a criterion based on the  \NUVrK\ diagram to separate the passive and star-forming  galaxies.
 To test the validity of the above color criterion, we show the distribution in the \NUVrK\ diagram of galaxies with morphological information provided by the  Zurich estimator of structural type (ZEST) catalog \citep{scarlata2007}. The ZEST classification uses three morphological classes: Early type, Disk and Irregular, with subclasses describing the degree of `irregularity' in the early-type class (i.e., 0 for regular, 1 for irregular), and the contribution of the bulge for disk galaxies (i.e., from 0 for bulge dominated galaxies, to 3 for pure, bulge-less disks). We also consider the ellipticity class for the galaxies classified as disks. This traces the galaxy inclination  with an ellipticity of 0 corresponding to a face-on galaxy  and up to 3 for an edge-on galaxy.
 %
%%%%%%%%%%%%%%%%%%%%%%%%%%%%%%
\begin{figure}
	\centering
\includegraphics[width=\hsize]{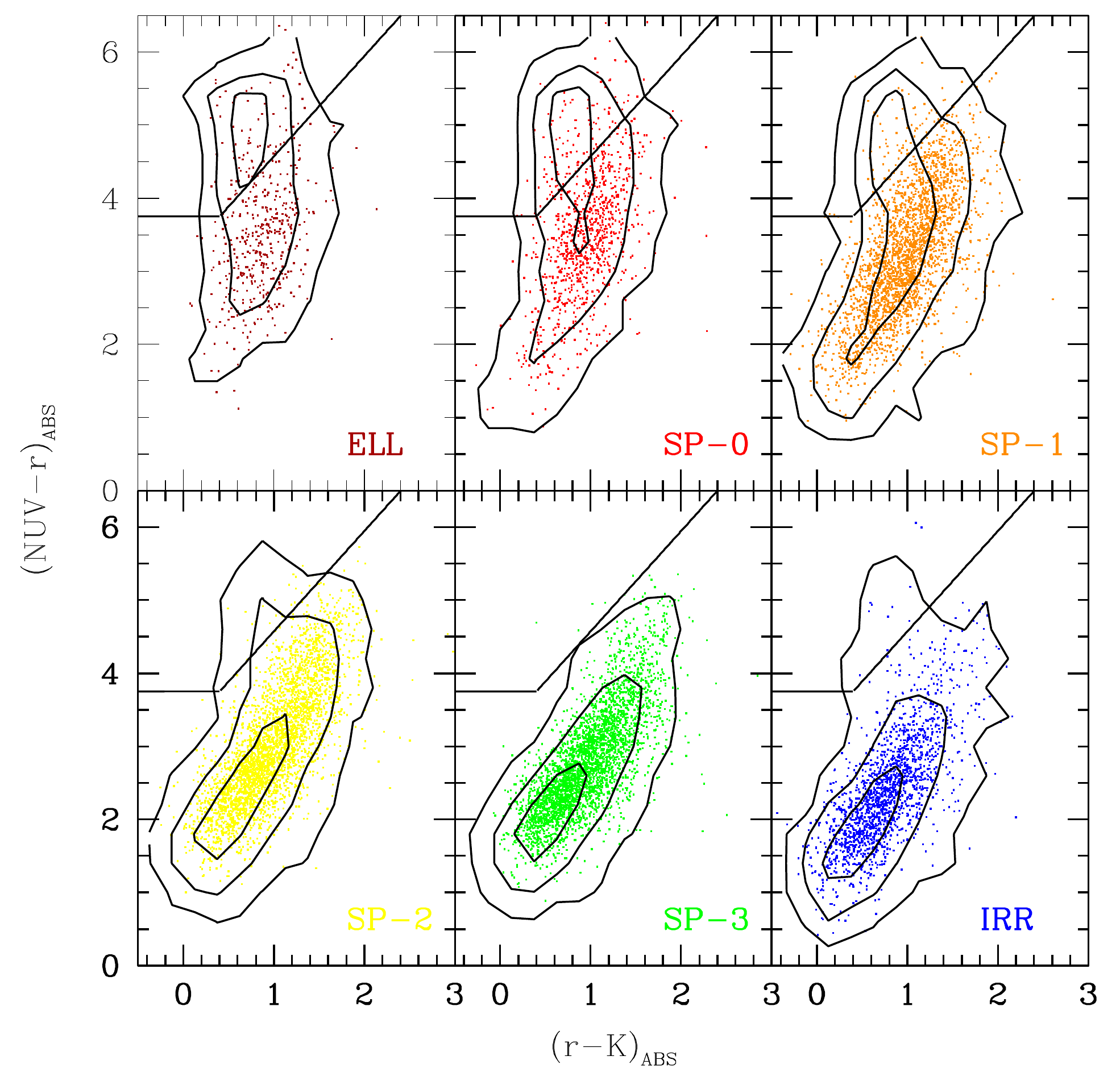}   
\caption[]{ Distribution of  the morphologically selected samples in the \NUVrK\ diagram.
 The density contours (1/2, 1/10,1/100 of the peak) refer to the whole morphological sample from \citet{scarlata2007}, while the dots  refer to the 24\,\micron-select  subsample.   
 }
\label{fig:morph}
\end{figure}
%%%%%%%%%%%%%%%%%%%%%%%%%%%%%%%%%%%%%%%%%%
%
  The ZEST catalog includes galaxies down to $I_{AB}\le 24$ and we restrict the sample to galaxies with good flags. The distributions of the different morphological classes in the \NUVrK\ diagram  are shown in Fig.~\ref{fig:morph} for the whole ZEST sample (as density contours) and for the 24\,\micron\ selected subsample (small dots).  
 We detect a clear evolutionary sequence in this diagram, with the Irregular (IRR) and Spiral-disk dominated (Sp-3, and  Sp-2) galaxies showing only blue colors, typical of active star-forming galaxies. The Spirals with a growing contribution of bulge (Sp-1 and Sp-0) and the
  early-types (ELL) show an increasing fraction of their population to lie in the passive region  (i.e., top-left side) of the diagram. The 24\,\micron\ sample tends to lie in the blue plums of the early type (ELL) and  bulge dominated spiral (Sp-0) samples.  The blue plum in the early type class could be due to some residual of star formation activities (the plum is present in the two subclasses based on the regularity criterion: ELL-0 and ELL-1). \\
 The Spiral disk-dominated  population (Sp-3, and  Sp-2)  extend to relatively red colors in $(NUV-r)$ and $(r-K)$ (top-right part of the diagram), where high values of IRX are also observed.
 As discussed by \citet{Patel2011}, in the UVJ diagram, the disk inclination can be indeed responsible for this extreme reddening.  This is indeed supported by Fig.~\ref{fig:ell}, where we show  the mean values of the ellipticity parameter for the disk population, in different redshift bins 
(see also the discussion in Section~\ref{sec:discu}).  
%
%%%%%%%%%%%%%%%%%%%%%%%%%%%%%%
\begin{figure}
	\centering
\resizebox{\hsize}{!}{\includegraphics{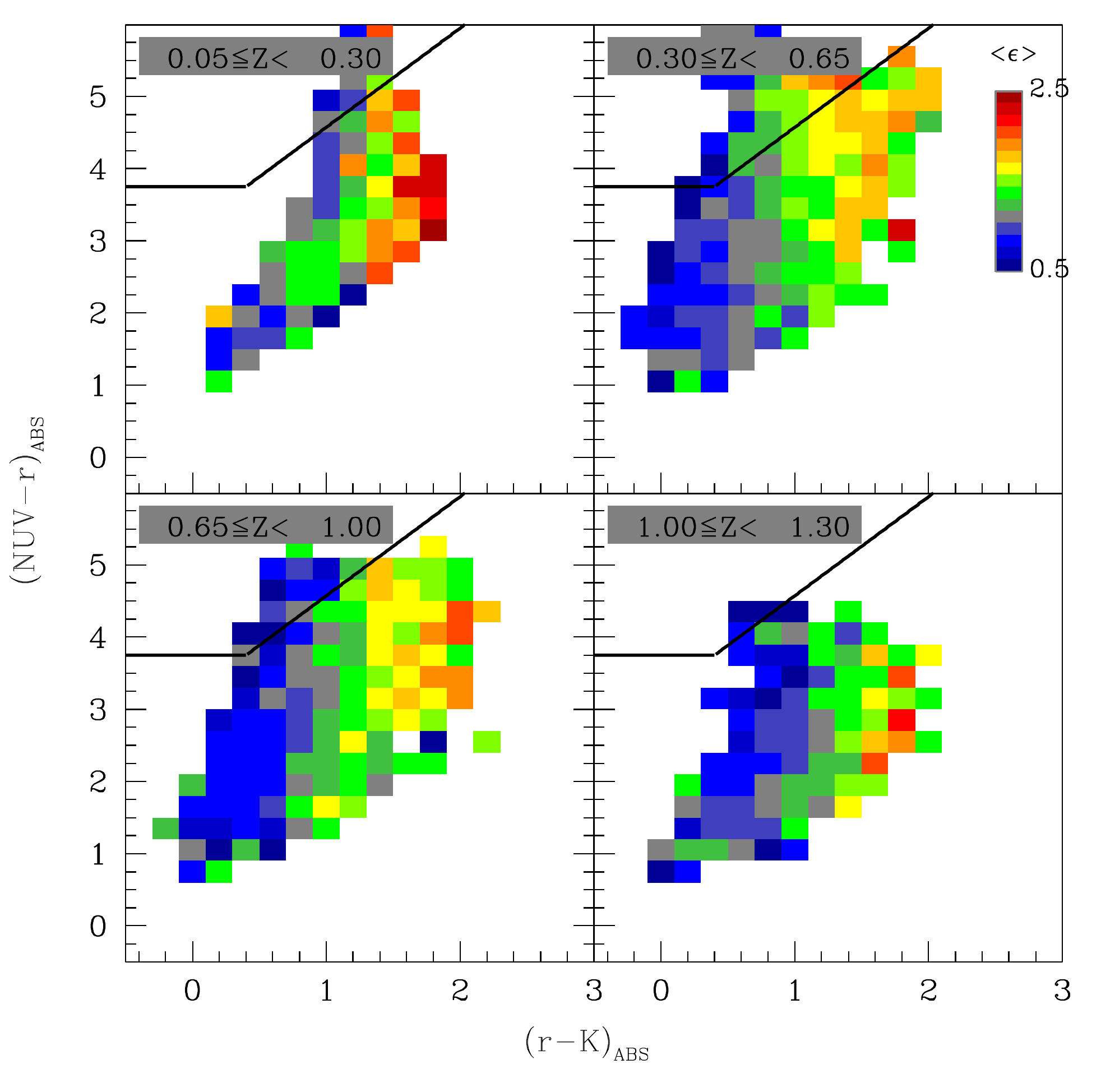}}   
\caption[]{ Mean value of the ellipticities (color coded) in the \NUVrK\ diagram for  the morphological sample  from \citet{scarlata2007}.  }
\label{fig:ell}
\end{figure}
%%%%%%%%%%%%%%%%%%%%%%%%%%%%%%%%%%%%%%%%%%
%
\section{The average spectral energy distributions along the vector NRK } 
\label{sec:sed}
%
%%%%%%%%%%%%%%%%%%%%%%%%%%
\begin{figure*}
\centering
\includegraphics[width=17cm]{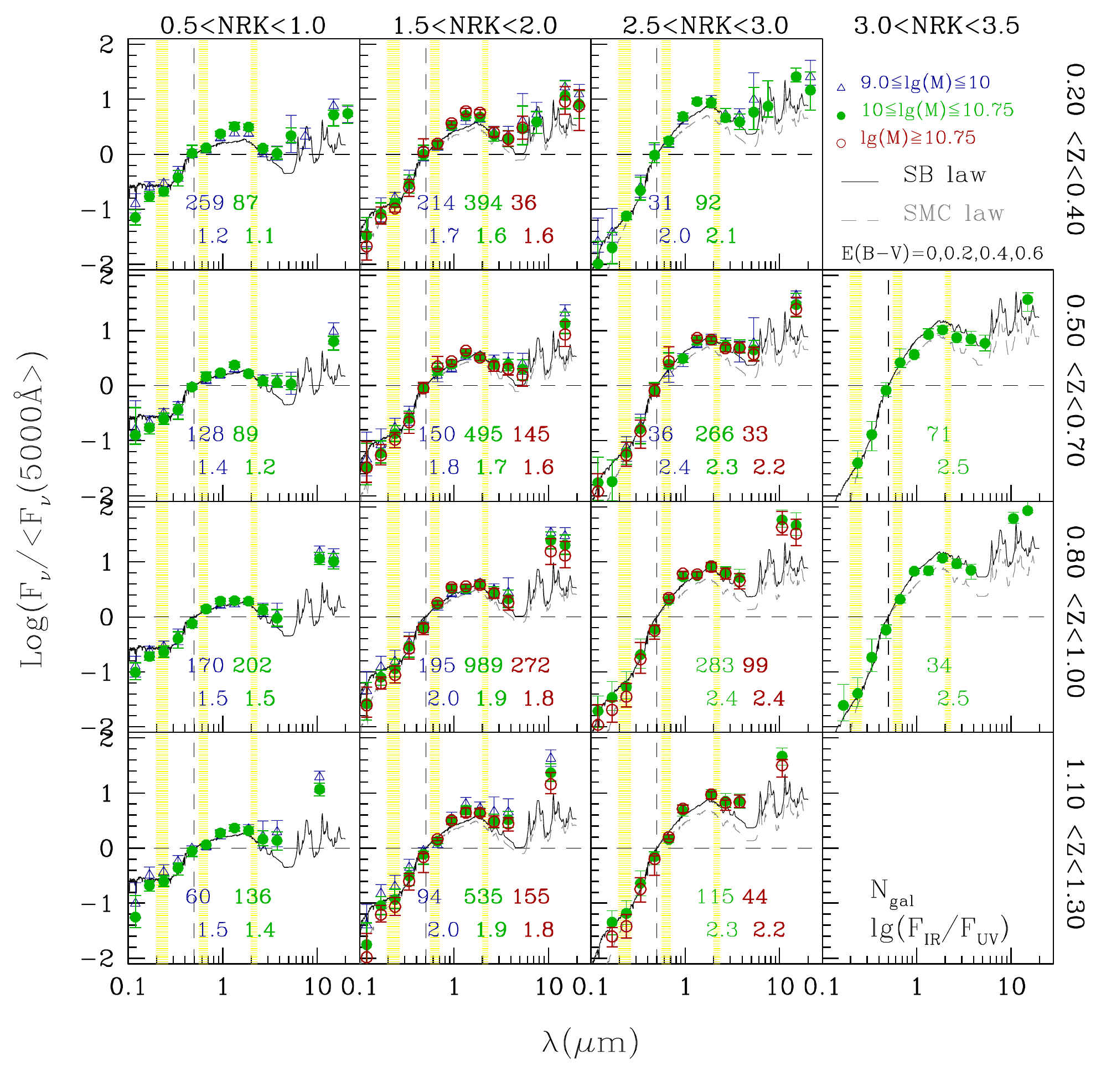}
\caption[]{ Evolution of the average rest-frame SEDs with \NRK\ for the 24\,\micron-selected sample. Each SED is reconstructed by using all the available COSMOS pass-bands. 
%The number of sources used to build each SED is reported in each panel, along with the ratio $\log[F_{24/(1+z)}/F_{2300}]$. 
Galaxies are divided in bins of \NRK\ (columns of the figure), redshift (rows) and stellar mass (blue, green and red symbols in each panel). 
The number of galaxies for each sample is shown in each panel. We also report  the logarithm of the mean ratio $F_{IR}/F_{UV}$  as an indicative value of the IRX.
 In the first bin of \NRK\ (i.e., first column of the figure) we overplot the dust-free template of young star-forming galaxy. In the other three bins of \NRK\ (i.e., three right-most columns) we overplot the same template, but with increasing reddening excess ($E(B-V)=0.2,\,0.4\,,0.6$, respectively), assuming the \citet{Calzetti2000} attenuation law (solid, black line) and the SMC extinction law (dashed, gray line). The three vertical bands  refer  to the  bandwidths of the NUV, r and K filters.  
  }
\label{fig:sed}
\end{figure*}
 The evolution of \irx\  with \NRKv\  vector  should be reflected in the shape of the galaxy SEDs
  when moving from low to high value of \NRK.  
 In Fig.~\ref{fig:sed}, we have reconstructed  the averaged rest-frame SEDs in different bins of \NRK, redshift and stellar mass, by using the 31 broad and medium bands available in the COSMOS catalog. They are reconstructed in logarithmic wavelength bins ($\Delta \log (\lambda)=$0.15).  Before computing the average and the dispersion in each wavelength bin, the low resolution spectrum of all the galaxies was first normalized at $\lambda\sim 0.5\micron$  (dashed lines).  In each panel, we show the number of galaxies used to reconstruct the averaged SEDs  and as an indicative value of the IRX, we show the 
 ratio $\log(F_{IR}/F_{NUV})$, where the $F_{IR}$ is the redshift corrected 24$\mu$m flux of the averaged SED ($F_{24\mu/(1+z)}$)  and $F_{NUV}$  the rest-frame flux at  0.23$\mu$m ($F_{0.23\mu}$).  As a reference, we overplot  the template of a young star-forming galaxy (SB2\footnote{We focus only on the UV to Mid-IR part of the template  with $\lambda\le 5\micron$} from \citealt{Ilbert2009}). For the first bin $0.5\le NRK\le 1$, corresponding to the bluest population in our sample, no attenuation is applied to the template. For the three other bins of \NRK, we increase the amount of dust reddening  to   $E(B-V)=0.2,0.4,0.6$,  assuming a starburst attenuation law (solid black lines) and a SMC extinction law (gray dashed lines).\\
  A strong evolution of the SED shape is observed with increasing  \NRK. We can  quantitatively reproduce this evolution by increasing the reddening excess applied to our SB2 star-forming template. On the other hand, redshift and stellar mass appear to play a secondary role in shaping the SED properties. 
  % Indeed a minor difference in SEDs is observed between the three stellar mass bins with the more massive ones showing a lower UV flux on average than less massive ones but this effect is accompanied by a lower Far-IR flux which in first order preserves the IRX ratio, as reported in each panel.
  This provides an additional evidence that \NRK\  is sensitive to the global budget between UV and Far-IR emission in galaxies, regardless of other properties.  It also supports our choice of neglecting the dependence on the stellar mass in Equation~\eqref{eq:irx}.
  However, in \citet{Lefloch2013} we  extend our analysis to a complete mass selected sample using the stacking technique at 24, 220, 350 and 500\,\micron\ and  we show that the  \irx\ vs \NRK\ relation should include a corrective term based on stellar mass when considering galaxies with stellar masses  $M_{\star} \le10^{9.3}\,\Msun$.
%
%%%%%%%%%%%%%%%%%%%%%%%%%%%%%%%%%%%%%%%%%%%%  
\section{Dependence with sSFR}
\label{app:dep_nrk}
\begin{figure}
\centering 
\includegraphics[width=\hsize]{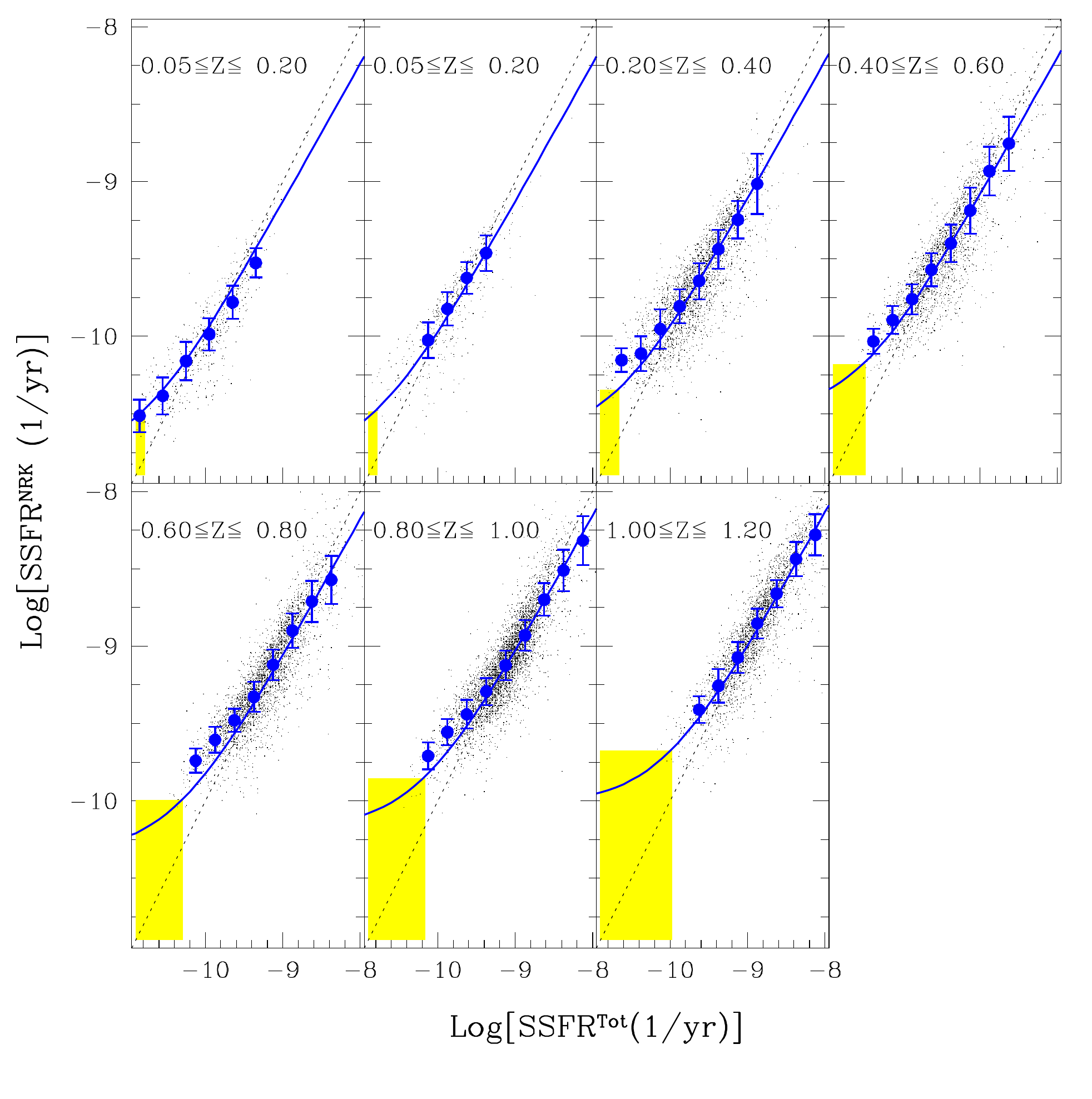}
\caption[]{The predicted specific SFR ($\sSFRnrk$) vs the  reference  sSFR  ($\sSFR_{tot}$)  for the star-forming population in the local SWIRE  (top left panel) and COSMOS (other panels) samples. The mean and the sigma per bin of $\sSFR_{tot}$  are shown as blue symbols and  the solid lines show the  analytical fit  described in the text. The yellow area corresponds to the limit where $\SFRnrk > 2 \times SFR_{tot}$. }
\label{fig:depnrk}
\end{figure}

 As observed in Fig.~\ref{fig:dirx1},  the difference between the IR luminosity derived from the 24\,\micron\ observations and the \NRK\ method varies with the specific SFR.  In particular, we have shown that the \NRK\ method over-estimates $\Lir$ for galaxies with low sSFR. 
  In Fig.~\ref{fig:depnrk} we compare, in different redshift bins, the predicted specific SFR ($\sSFRnrk$) with the reference sSFR (derived with the 24\,\micron , $\sSFR_{tot}$), with both sSFR estimated with Equation~\eqref{eq:sfr}. 
%%%%%%%%%%%%%%%%%%%%%%%
%%%%%%%%%%%%%%%%%%%%%%%%%%%%%%%%
%
The $\sSFRnrk$ tends to saturate while the true sSFR keeps declining toward galaxies with low star formation activity, and this deviation varies with redshift. We propose an analytical fit which allows us to reproduce this deviation over the entire redshift range: 
\begin{equation}
\log(\sSFRnrk^{COR})= -9.5 +1.1 [ \log(10^{y_0}  - 10^{a(z)}) - b(z) ]
\end{equation}
where, $y_0=\log(\sSFRnrk)+9.5$,  $a(z)=-1.4+0.8\ z$ and $b(z)=-0.1\ (1-z)$ and z is the redshift. By mean of this equation, we can define a sSFR threshold below which the \NRK\ method over-estimates the SFR by a given factor. By choosing a factor 2, the redshift evolution of this threshold can be simply described by $\sSFRnrk_{lim}(z)=-10.6 + 0.8\ z$. These thresholds are shown as shaded yellow regions in Fig.~\ref{fig:depnrk}.  

\end{document}